\documentclass[aps,groupedaddress,twocolumn,showpacs,longbibliography,10pt]{revtex4-1}
\usepackage{amsmath}
\usepackage{amsthm, amssymb,bbold}    
\usepackage{braket}              
\usepackage{graphicx}
\usepackage{color}
\usepackage{times, verbatim}      
\usepackage{mathtools,tikz}
\usetikzlibrary{matrix}
\usepackage{natbib}
\usepackage{multirow}
\usepackage{hyperref}
\usepackage{wrapfig,lipsum,booktabs}
\hypersetup{
       colorlinks=true,
}

\begin{document}

\title{
Group theoretic approach to many-body scar states in fermionic lattice models
}

\author{K. Pakrouski,$^{1,2}$ P.N. Pallegar,$^{2}$  F.K. Popov,$^{2,3}$ I.R. Klebanov$^{2,4,5}$         }
\affiliation{$^{1}$Institute for Theoretical Physics, ETH Zurich, 8093 Zurich, Switzerland}
\affiliation{$^{2}$Department of Physics, Princeton University, Princeton, NJ 08544, USA}
\affiliation{$^{3}$Department of Physics, New York University, New York, NY 10003, USA}
\affiliation{$^{4}$Princeton Center for Theoretical Science, Princeton University, Princeton, NJ 08544, USA}
\affiliation{$^{5}$Institute for Advanced Study, Princeton, NJ 08540, USA}

\begin{abstract}
It has been shown \cite{pakrouski2020GroupInvariantScars} that three families of highly symmetric states are many-body scars for any spin-1/2 fermionic Hamiltonian of the form $H_0+OT$, where $T$ is a generator of an appropriate Lie group. One of these families consists of the well-known 
$\eta$-pairing states.
In addition to having the usual properties of scars, these families of states are insensitive to electromagnetic noise and have advantages for storing and processing quantum information. In this paper we show that a number of well-known coupling terms, such as the Hubbard and the Heisenberg interactions, and the Hamiltonians containing them, are of the required form and support these states as scars without fine-tuning. The explicit $H_0+OT$ decomposition for a number of most commonly used models, including topological ones, is provided. To facilitate possible experimental implementations, we discuss the conditions for the low-energy subspace of these models to be comprised solely of scars. Further, we write down all the generators $T$ that can be used as building blocks for designing new models with scars, most interestingly including the spin-orbit coupled hopping and superconducting pairing terms. We expand this framework to the non-Hermitian open systems and demonstrate that for them the scar subspace continues to undergo coherent time evolution and exhibit the "revivals". A full numerical study of an extended 2D $tJU$ model explicitly illustrates the novel properties of the invariant scars and supports our findings. 
 \end{abstract}

\date{\today}

\pacs{}

\maketitle

\section{Introduction}

In recent literature there has been considerable interest in the many-body scar states
 ~\cite{RydbergExperimentRevivals,AbaninScarsSU2Dynamics,Shiraishi2017ScarsConstruction,Moudgalya:2018,Turner_2018,Khemani:2019vor,Sala_2020,Prem:2018,Schecter:2019oej,SciPostPhys.3.6.043,IadecolaHubbardAlmostNUPRL2019,Shibata:2020yek,michailidis2020stabilizing,2020MarkMotrEtaPairHub,PRLPapicClockModels,VedikaScarsVsIntegr,Pal2020ScarsFromFrustration,mark2020unified,iadecola2020quantum,moudgalya2020etapairing,Magnifico:2019kyj,pakrouski2020GroupInvariantScars,2020arXiv200710380R,ODea:2020ooe,Nielsen2020ChiralScars,Hsieh2020PXP2D,Regnault2020MPStoFindSc,FloquetPXPScars,PapicWeaklyBrokenAlgebra,kuno2021multiple,banerjee2020quantum,
chaoticDickeAllScarred2021,maskara2021DrivenScars,langlett2021rainbow,2021arXiv210807817R,2021arXiv211011448T,Schindler:2021lma} (for their pedagogical overviews, see \cite{Serbyn:2020wys,Moudgalya:2021xlu}). 
The scar states are typically spread throughout the energy range of the spectrum and are, therefore, relevant when the excitation energy is high. 
They do not obey the eigenstate thermalization hypothesis (ETH) \cite{deutsch1991quantum,srednicki1994chaos,rigol2008thermalization}, one of the most fundamental conjectures that allows us to bridge quantum mechanics with statistical physics. On the practical side, an initial state made of many-body scar states repeatedly returns to itself in time evolution preventing the loss of quantum information through thermalization. This offers intriguing prospects for quantum computing.

Examples of many-body scars have been found in a number of systems including correlated electron models with the Heisenberg \cite{KatsuraScarsInSpinChains,SchoutensOnsagerScarsProperties,RichterScarsFromFrustration,mark2020unified}, Hubbard \cite{moudgalya2020etapairing,mark2020unified,IadecolaHubbardAlmostNUPRL2019,tilted1DHubbardPapic} and density-density \cite{HatsugaiFlatBandScar,DeTomasiScarsWithDensityDensity} interactions. These findings represent a scattered puzzle of scar phenomenology lacking a clear fundamental mechanism.

In our previous paper \cite{pakrouski2020GroupInvariantScars} (see also \cite{2020arXiv200710380R,ODea:2020ooe}), we presented a general strategy for systematically designing the Hamiltonians with a many-body scar subspace 
$\mathbb{S}$ invariant under the action of a continuous group $G$, which is bigger than the symmetry group of the Hamiltonian. 
The general form of such Hamiltonians is
$H= H_0+ \sum_j O_j T^G_j$, where $T^G_j$ are generators of the symmetry group $G$ and $O_j$ are a set of operators such that the product $O_j T^G_j$ is Hermitian. 
$H_0$ must admit the states in $\mathbb{S}$ as eigenstates and the revivals are observed when the gaps between the corresponding eigenvalues have a common divisor.

In this work, we demonstrate that many of the commonly used condensed matter models of interacting electrons actually happen to be of this form and specify their group-invariant scar subspaces $\mathbb{S}$. This applies to the Hubbard, Heisenberg, and some other interactions, and any models constructed out of them on various lattices and in arbitrary dimension such as the extended 2D $tJU$ model that we consider in detail as a prototype. Most of these models can be readily implemented in experiment paving the way for their systematic experimental studies. Our results show that some of the known scar states \cite{moudgalya2020etapairing,2020MarkMotrEtaPairHub} can indeed be explained by the mechanism of Ref. \cite{pakrouski2020GroupInvariantScars}. An important feature of our construction is that $G$ is a large Lie group, such as O$(N)$ or SU$(N)$, whose rank is of order $N$, the number of lattice sites. The generators of such groups include the nearest neighbor hopping terms, and using them we can construct a variety of (approximately) local lattice models. 

We also extend the framework of \cite{pakrouski2020GroupInvariantScars} to non-Hermitian Hamiltonians. 
In particular, we study a model where the product $OT$ is not Hermitian. This opens the way to studying the coherent time evolution of a group-invariant scar subspace in open systems. To our knowledge, many-body scars at finite energy density in non-Hermitian systems have not been discussed previously \footnote{A phenomenon equivalent to scars at zero energy density was mentioned in the Supplementary material of Ref. \cite{BucaNature2019}.}, while the non-stationary, periodic phenomena in dissipative strongly-interacting systems is an emergent hot topic \cite{bookerHubbardPlusDissipation,Oberreiter2021PRL,Ikeda2020PRL,Song2020PRBDynamicalEta,Soriente2021,buca2021algebraic}.

In Sec. \ref{sec:nOnU} we specify the three families of scar states that are relevant to our discussion of spin-$\frac12$ fermionic models. We describe their structures, and the generators $T$ annihilating them. These generators may be viewed as the free Hamiltonians. 
In Sec. \ref{sec:interTerms} we show how the well-known interaction terms, such as the Hubbard and Heisenberg interactions, decompose in terms of the same generators. As a consequence, the models including any linear combinations of such interaction terms also have $H_0+OT$ decomposition. In Table \ref{tab:knownModelsAsOT} we give these decompositions for several models, such as the $J_1-J_2$ and Haldane-Hubbard models. We conclude with an example (Sec. \ref{sec:tJUNumerics} ) where the full construction is detailed for the $tJU$ model and the numerical evidence of many-body scar states is provided.

\section{Group generators and invariant states \label{sec:nOnU}}

  \begin{table}[t!]
  	\centering
  	\caption{ Properties of the invariant subspaces with respect to the groups acting on the lattice index and spin and pseudo-spin groups. 
  }	
  	\label{tab:Reps}
	
  		\begin{tabular}{|c|c|c|c|}
  			\hline
  								 states	& lattice index							& spin					& pseudo-spin	\\
	 						\hline
	 					$\ket{n^\eta}$ 		& $ \widetilde{\rm U}(N)$-invariant					&singlet					& spin-$\frac{N}{2}$ 				\\
	 		\hline
	 					$\ket{n^\zeta}$		& U$(N)$-invariant							& spin-$\frac{N}{2}$			& singlet 				\\
  			\hline
	 					$\ket{n^\eta}'$   	& $\widetilde{\rm U}(N)'$-invariant					& singlet					& spin-$\frac{N}{2}$				\\
	 		\hline
  		\end{tabular}
	
  \end{table}

Consider a lattice of $N$ sites with each one hosting a complex spin-1/2 fermion (electron). Their two spin states are created on site $j$ by operators $c^\dagger_{j\sigma}$, where
$\sigma=\uparrow, \downarrow$. We do not assume any spatial structure in general but will specify it where necessary. 
It is useful to think of the Hilbert space of $2N$ fermionic degrees of freedom as factorized according to the representations of one of the following groups: 
\begin{gather}
G_1 = {\rm O} (N)\times {\rm O}(4) = {\rm O}(N)  \times {\rm SU}(2)_{spin} \times {\rm SU}(2)_\eta \ ,
\nonumber \\
G_2 = {\rm U}(N)  \times {\rm SU}(2)_{spin}= {\rm U}(1)_Q \times {\rm SU}(N)  \times {\rm SU}(2)_{spin}\ ,\nonumber \\ 
G_3 = \widetilde {\rm U}(N)  \times {\rm SU}(2)_{\eta}= {\rm U}(1)_M \times \widetilde {\rm SU}(N)  \times {\rm SU}(2)_{\eta}\ ,
\end{gather}
where the large rank groups O$(N)$, SU$(N)$, $\widetilde {\rm SU}(N)$ act on the site index. In particular, the factorization under ${\rm O}(N)\times {\rm O}(4)$ is natural if one works in the representation of $4N$ Majorana fermions \cite{Gaitan:2020zbm,Klebanov:2018nfp}.
Due to the peculiar nature of the Hilbert space, the representations of a group acting on the site index are locked to particular representations of the dual 
${\rm O}(4)={\rm SU}(2)_{spin} \times {\rm SU}(2)_\eta$ \cite{Gaitan:2020zbm} (the ${\rm SU}(2)_\eta$ group is often called the pseudospin).

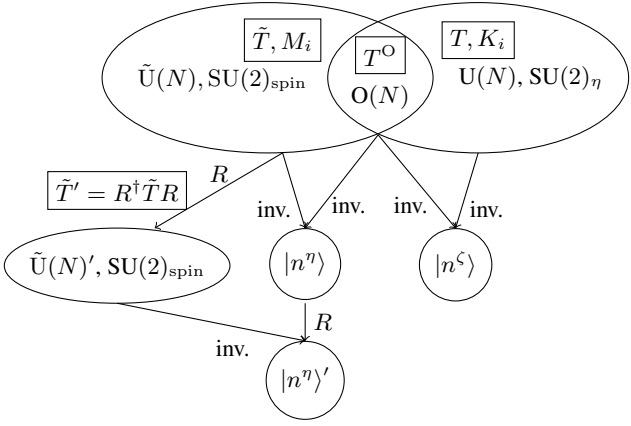
\begin{figure}
	\centering
	\begin{tikzpicture}
		\draw (1.3,0) ellipse (2cm and 1cm);
		\draw (-1.5,0) ellipse (2.2cm and 1cm);
		\node at (0,-0.25) {O$(N)$};
		\node at (-1.3,0.5) [draw] {$\tilde{T},M_i$};
		\node at (1.3,0.5) [draw] {$T,K_i$};
		\node at (0,0.3) [draw] {$T^{\rm O}$};
		\node at (-2.1,0) {$\tilde{{\rm U}}(N), {\rm SU}(2)_{\rm spin}$};
		\node at (2,0) {U$(N)$, SU$(2)_{\rm \eta}$};
		\node at (-1,-2)[draw,circle, anchor=north] {$\ket{n^\eta}$};
		\draw[->] (-1.3,-1)--node[pos=0.75, left] {\small inv.} (-1,-2);
		\node at (1,-2)[draw,circle, anchor=north] {$\ket{n^\zeta}$};
		\draw[->] (1.3,-1)-- node[pos=0.75, right] {\small inv.} (1,-2);
		\draw[->]  (-0.03,-0.76) -- node[pos=0.75, right] {\small inv.} (-1,-2);
		\draw[->]  (-0.03,-0.76) -- node[pos=0.75, left] {\small inv.} (1,-2);
		\draw[->]   (-1.3,-1)--  node[pos=0.5, above] {$R$} (-3,-2);x
		\draw (-3.5,-2.5) ellipse (1.5cm and 0.5cm);
		\node at (-3.5,-2.5) {$\tilde{{\rm U}}(N)'$, SU$(2)_{\rm spin}$};	
		\node at (-3.5,-1.5) [draw] {$\tilde{T}'=R^\dagger\tilde{T}R$};
		\draw[->] (-3.5,-3)-- node[pos=0.75, left, anchor=north east] {\small inv.} (-1,-3.5);
		\node at (-1,-3.5)[draw,circle, anchor=north] {$\ket{n^\eta}'$};
		\draw[->] (-1,-3)-- node[pos=0.5, right] {$R$}(-1,-3.5);
	\end{tikzpicture}
	\caption{Interrelations between the three invariant subspaces and the corresponding groups. The three groups that act naturally on the Hilbert space are shown inside the ellipses, the corresponding generators are shown inside the rectangular boxes, and the invariant states are shown inside the circles.} 	
	\label{fig:invSubSpStruct}
\end{figure}

The symmetry properties of the invariant states we are going to consider in this work are summarised in Table \ref{tab:Reps} (the states are given explicitly
in \eqref{stateNu},\eqref{eq:statenO},\eqref{eq:etaPairState}). Note that each of these subspaces is invariant under two groups, a large group acting on the site index and a small group acting on the spin or pseudospin. For example, the states $\ket{n^\zeta}$ are invariant under both U$(N)$ and SU$(2)_\eta$. Therefore, the generators of each of these groups 
$T_{{\rm U}(N)}$ and $T_{{\rm SU}(2)_\eta}$ and the commutators $[T_{{\rm U}(N)}, T_{{\rm SU}(2)_\eta}]$ all annihilate \footnote{some of these terms involve a sum over all system sites and should be excluded if a purely local Hamiltonian is desired} the $\ket{n^\zeta}$ states and can be used as $T$ in the $H_0+OT$ Hamiltonian.
Further insights into the interrelations between the various groups and the invariant states are provided in Fig. \ref{fig:invSubSpStruct} and Appendix \ref{sec:AppSemi}.

In order to construct the Hamiltonians such that the invariant states are many-body scars, we now discuss the generators of all the relevant groups.
We will use the standard notation for the fermion numbers at site $i$: 
\begin{gather}
n_{i\uparrow} = c^\dagger_{i\uparrow} c_{i\uparrow}\ , \quad  n_{i\downarrow} = c^\dagger_{i\downarrow} c_{i\downarrow}\ , \quad
n_i = n_{i\uparrow} + n_{i\downarrow} \ .
\end{gather}
The total fermion number is 
\begin{gather}
Q = \sum\limits^N_{i=1} n_i
\ ,
\end{gather}
and we will call the corresponding symmetry U$(1)_Q$ (the actual generator is $(Q-N)/2$). 

The generators of the rotation group SU$(2)_{\rm spin}$ are given by 
\begin{gather} \label{hop:zeta}
Q_A= \sum_{i=1}^N S^A_i\ ,  \qquad A=1,2,3\ .
\end{gather}
The spin operator at site $i$ is
\begin{gather}
\label{eq:SiInFerm}
S^A_i=  \frac{1}{2} \sum_{\alpha,\beta} c^\dagger_{i \alpha} \sigma^A_{\alpha \beta} c_{i \beta}\ ,
\end{gather}
where $\sigma^A$ are the Pauli matrices, and the Greek indices take two values, $\uparrow$ and $\downarrow$. In particular, 
\begin{gather}
2 S^3_i=M_i= n_{i\uparrow} - n_{i\downarrow}
\end{gather}
is the magnetization at site $i$. The symmetry corresponding to total magnetization is U$(1)_M\subset {\rm SU}(2)_{\rm spin}$.
  
Another important group is the pseudospin \cite{etaPairingYang89,yang1990so,ZhangHubbardSO41991}, which is denoted by SU$(2)_{\eta}$. Its generators are
\begin{gather}\label{hop:eta}
	\eta^+ = \sum_j c^\dagger_{j\uparrow}c^\dagger_{j\downarrow} =  \frac{1}{2} \sum_{j, \sigma, \sigma'} c^\dagger_{j\sigma} c^\dagger_{j\sigma'} \epsilon_{\sigma\sigma'}\ , \\ \notag
	\eta^-=(\eta^+)^\dagger,\quad \eta^3=\frac{1}{2} (Q-N) \ ,
\end{gather}
so that U$(1)_Q \subset {\rm SU}(2)_{\eta}$.

The generators of SU$(2)_{\rm spin}$ and SU$(2)_{\eta}$ involve summation over all lattice sites. Therefore, using such generators in the interaction terms $\sum_j O_j T^G_j$ will produce a very non-local Hamiltonian. Instead, we will follow the suggestion in our previous paper \cite{pakrouski2020GroupInvariantScars} to construct these interaction terms using the generators of large rank groups, such as SU$(N)$, which include the nearest-neighbor spin-independent hopping terms.

The full set of generators of SU$(N)$ are the hopping terms with generally complex amplitudes: 
\begin{gather} \label{hop:T}
T_{ij} = \lambda \sum_\sigma c^\dagger_{i\sigma} c_{j\sigma} + {\rm h.c.} ,\quad \lambda \in \mathbb{C}.
\end{gather}
There is an SO$(N)$ subgroup of SU$(N)$ whose generators are the spin-preserving hopping terms with imaginary amplitudes
\begin{gather} \label{hop:tilT}
T^{O}_{kl} = 
 i \sum_\sigma (c^\dagger_{k\sigma}  c_{l\sigma}- c^\dagger_{l\sigma}  c_{k\sigma}) \ .
\end{gather}
They are invariant under ${\rm SU}(2)_{spin} \times {\rm SU}(2)_\eta$.

An alternate basis for the generators of SU$(N)$ is
\begin{gather} \label{altbas}
T^a = \sum_{i,j, \sigma} t^a_{ij} c^\dagger_{i\sigma} c_{j\sigma}\ , \qquad a=1, \ldots N^2-1\ ,
\end{gather}
where $t^a$ are the traceless Hermitian $N\times N$ matrices.
The simple root generators of this algebra could be chosen as the nearest-neighbor hoppings
\begin{gather}
T_{i} = \sum_\sigma c^\dagger_{i \sigma} c_{n(i)\sigma}, \label{eq:simroot}
\end{gather}
where $n(i)$ is a nearest neighbour of the site $i$ on a given lattice. Commuting simple roots we can restore the whole algebra \eqref{hop:T}. Any Hermitian spin-independent hopping terms on any lattice  
are linear combinations of $T_{ij}$ \eqref{hop:T}. They belong to the SU$(N)$ algebra and annihilate the SU$(N)$ singlets (and consequently also the singlets of the larger group U$(N)$).

There is another, less obvious, group $\widetilde {\rm SU} (N)$, whose generators are \eqref{hop:tilT}
and the spin-orbit coupled hopping 
\begin{gather}
\tilde{T}^{\rm sym}_{kl} = \sum_{\alpha, \beta}
\left ( c^\dagger_{k\alpha}\sigma^2_{\alpha\beta}  c_{l\beta}+ c^\dagger_{l\alpha}\sigma^2_{\alpha\beta}  c_{k\beta}\right ) \ .
\label{newgen}
\end{gather}
The spin-preserving hopping terms \eqref{hop:tilT} are generators of the SO$(N)$ which is a subgroup of both $\widetilde {\rm SU} (N)$ and SU$(N)$ (see Fig. \ref{fig:invSubSpStruct}).

By combining operators $\tilde{T}_{kl}$ \eqref{newgen} and $Q_A$ \eqref{hop:zeta} we obtain more general spin-orbit coupled hopping terms
\begin{gather} \label{hop:Ttilalpha}
\tilde{T}^{A}_{ij}=\sum_{\alpha, \beta} \left ( c_{i \alpha}^\dagger \sigma^A_{\alpha \beta} c_{j \beta} +  c_{j \alpha}^\dagger \sigma^A_{\alpha \beta} c_{i \beta} \right )
\end{gather}
that annihilate the singlets of $\widetilde {U} (N)$ and SU$(2)_{spin}$ ($\ket{n^\eta}$ states).
A special case of these generators is the local spin operator $S^A_i $.

By commuting $\eta^+$ \eqref{hop:eta} and $T_{ij}$ \eqref{hop:T} we can get the operators that 
annihilate the singlets of SU$(2)_{\eta}$ and of U$(N)$ ($\ket{n^\zeta}$ states)
\begin{gather}\label{hop:T+}
	T^+_{ij} = \left[\eta^+,T_{ij}\right] = \lambda \sum_{\sigma, \sigma'} \epsilon_{\sigma\sigma'} c^\dagger_{i \sigma}  c^\dagger_{j \sigma'}, \qquad T^- = (T^+)^\dagger\ .
\end{gather}

There is a particular set of generators of U$(N)$ that we will find useful:
\begin{gather}
\label{eq:Ki}
K_i = n_i -1\ , \qquad i=1, \ldots N\ .
\end{gather}
For example,
\begin{gather}
K_N= n_N-1 = \frac{\eta^3}{N} - T^{N^2-1}
\ ,
\end{gather} 
where 
$T^{N^2-1}$ is the generator of SU$(N)$ corresponding to the matrix 
\begin{gather}
t^{N^2-1}= {\rm diag}\left (\frac{1}{N}, \frac{1}{N}, \ldots, -1+ \frac{1}{N}\right )\ .
\end{gather}
We have $\sum_{i=1}^N K_i = 2\eta^3$, and
the square of $K_i$ can be expressed in terms of the local magnetization: 
\begin{gather}
\label{eq:KiThroughMi}
K_i^2=1-M_i^2\ ,
\end{gather}
where we used $n^2_{i\uparrow}= n_{i\uparrow}$ and  $n^2_{i\downarrow}= n_{i\downarrow}$. 

\begin{figure}
	\centering
	\begin{tikzpicture}[baseline={(0,-4.5)}]
		\foreach \i in {1,...,8}
{
        \node[circle](\i) at (1*\i-1,0) {\i};
}

		\foreach \i in {9,...,16}
{
        \node[circle]
        (\i) at (16-1*\i,-1) {\i};
}

	\foreach \i in {17,...,24}
{
        \node[circle]
        (\i) at (1*\i-17,-2) {\i};
}
	\foreach \i in {25,...,32}
{
        \node[circle]
        (\i) at (32-1*\i,-3) {\i};
}
	\foreach \i in {1,3,...,31}
{
	\draw[fill=red] (\i) circle [radius=0.2];
}

\foreach \i in {2,4,...,32}
{
	\draw[fill=blue] (\i) circle [radius=0.2];
}

		\foreach \i in {1,...,8}
{
        \node[circle]
        (\i) at (1*\i-1,0) {\i};
}

		\foreach \i in {9,...,16}
{
        \node[circle]
        (\i) at (16-1*\i,-1) {\i};
}

	\foreach \i in {17,...,24}
{
        \node[circle]
        (\i) at (1*\i-17,-2) {\i};
}
	\foreach \i in {25,...,32}
{
        \node[circle]
        (\i) at (32-1*\i,-3) {\i};
}

	\draw[pink] (15)--(2);
	\draw[pink] (15)--(14);
	\draw[pink] (15)--(16);
	\draw[pink] (15)--(18);
	\draw[pink] (15)..controls (2.5,-1.65)..(12);
	\draw[pink] (15)..controls (3.5,-0.15)..(10);
	\draw[pink] (15)--(32);
	\draw[pink] (15)--(30);
	\draw[green] (23)--(11);
	\draw[green] (23)--(9);
	\draw[green] (23)--(27);
	\draw[green] (23)--(25);
	\draw[green] (23)..controls (4,-1.35)..(19);
	\draw[green] (23)..controls (5,-2.65)..(21);
	\end{tikzpicture}
	\hspace{0.5cm}
	
	\caption{A depiction of the bipartite square 2D lattice. The red lines indicate the ``even" hoppings that connect sites on different sublattices. The green lines indicate the ``odd" hoppings that connect sites on the same sublattice. The numbers labelling the sites show the lattice site index, which is acted on by one of the large groups such as SU$(N)$.}
		\label{fig:snake}
\end{figure}
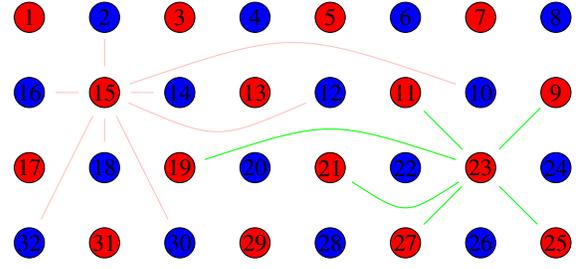

Now let us consider a special case of a bipartite lattice where the vertices are divided into two non-intersecting sets, which we can call red and blue (see Fig. \ref{fig:snake}). Here, in addition to the above operators, we can define the transformed generators 
\begin{gather}
\label{tildeprime}
\tilde{T}'=R^\dagger \tilde{T} R
\end{gather}
where
\begin{gather}
\label{eq:Rtransform}
R = i^{\sum\limits_{j\in B} n_j}\ , \quad R^\dagger R =I\ ,
\end{gather}
and the sum runs over the blue sites. The generators \eqref{tildeprime} 
form a group that we denote as $\widetilde{\rm SU}(N)'$.
A particular subset of the generators $\tilde{T}'$ is the nearest-neighbor hopping terms with a real coefficient
\begin{gather} \label{hop:T'}
T'_{\langle i, j \rangle} = \sum_\sigma c^\dagger_{i\sigma} c_{j \sigma} + h.c.
\end{gather}
Other ways of constructing a $\widetilde{\rm SU}(N)'$ algebra of longer-range hopping terms are discussed in the next sub-section. We will denote the spin preserving subgroup of this group as 
SO$(N)' \subset \widetilde{\rm SU}(N)'$.

We now provide the basis for each invariant subspace listed in the Table \ref{tab:Reps}. Each subspace should be invariant under the action of $H_0$
for the invariant states to be scars  \cite{pakrouski2020GroupInvariantScars}. A particular $H_0$ may
have eigenstates that are different from the basis discussed below but may be obtained from it by a rotation.

The subspace invariant under U$(N)$ and SU$(2)_\eta$ is spanned by $\ket{n^\zeta}$ which have the highest possible physical spin; namely, they form the spin-$\left(\frac{N}{2}\right)$ representation of SU$(2)_{spin}$ \cite{pakrouski2020GroupInvariantScars}: 
\begin{gather}
\ket{n^\zeta} = \frac{\zeta^n}{ \sqrt{\frac{N! n!}{(N-n)!}}} \ket{0^\zeta}\ , \qquad \ket{0^\zeta} =\prod_{j=1}^N c^\dagger_{j\downarrow}  \ket{0}\ , \label{stateNu}
\end{gather}
where $n=0, \ldots, N$, and 
\begin{gather}
\label{eq:zeta}
\zeta= Q_1 + i Q_2 = \sum_{j=1}^N c^\dagger_{j \uparrow} c_{j \downarrow}
\end{gather}
is the spin raising operator.

The states $\ket{n^\eta}$ are invariant under $\widetilde{\rm U} (N)$ and SU$(2)_{\rm spin}$; they form 
the $N+1$ dimensional representation of pseudospin SU$(2)_{\eta}$:
\begin{gather}
\ket{n^\eta} = \frac{(\eta)^n}{\sqrt{\frac{N! n!}{(N-n)!}}} \ket{0}\ , \qquad n=0, \ldots, N \ ,
\label{eq:statenO}
\end{gather}
where $\eta=\eta^+$ \eqref{hop:eta}. 
On a bipartite lattice we can further define the states $\ket{n^\eta}'$ that are invariant under
$\widetilde{\rm U} (N)'$ and form the $N+1$ dimensional representation of SU$(2)'_{\eta}$:
\begin{gather}
\ket{n^\eta}' = R\ket{n^\eta}=\frac{(\eta')^n}{\sqrt{\frac{N! n!}{(N-n)!}}} \ket{0} \  , \quad  \eta' = \sum_{j=1}^N e^{i\pi j} c^\dagger_{j\uparrow}c^\dagger_{j\downarrow}\ .
\label{eq:etaPairState}
\end{gather}
These states are known as the $\eta$-pairing states \cite{yang1962concept,etaPairingYang89} \footnote{In \cite{pakrouski2020GroupInvariantScars} these states were denoted as $\ket{n'_O}$.}. It is not hard to check that 
\begin{gather}
[S_i^A, \eta]= [S_i^A, \eta']= 0\ .
\label{eq:localspincomm}
\end{gather}
Combined with the fact that both families of states include the vacuum, that is annihilated by $S_i^A$, this means that all the $\ket{n^\eta}$ and $\ket{n^\eta}'$ states are annihilated by the local spin operators $S_i^A$.

The importance of the spin and pseudopspin SU$(2)$ groups has been previously understood in the context of the Hubbard model \cite{etaPairingYang89,yang1990so,ZhangHubbardSO41991}. 
These two SU$(2)$ groups are related by the Shiba transformation, that acts on the fermion operators in the following way:
\begin{gather}
c_{j\uparrow} \to c_{j\uparrow}\ ,\quad  
c_{j\downarrow} \leftrightarrow (-1)^j c^\dagger_{j\downarrow}\ .
\end{gather}  
 We note that this transformation interchanges the $\ket{n^\eta}'$  and $\ket{n^\zeta}$ states and leaves the SO$(N)$ generators \eqref{hop:tilT} invariant. The Hubbard Hamiltonian is also known \cite{hubbard1DbookShiba} to be invariant under the Shiba transformation.

All three families of scars have a very high degree of symmetry with respect to the spatial index $j$. As a consequence, when the sites are arranged into a lattice, all three families are translation and permutation-invariant. In Table \ref{tab:stAnnihilators} we summarize the action of all the generators we mentioned on the three families of the invariant states. This table provides a simple way of constructing Hamiltonians with one of the families as scars  - any linear combination of the operators with 0 in a corresponding row can be used as an operator $T$ in the general $H_0+OT$ form. 
In Sec. \ref{sec:interTerms} we show that many of the commonly known interaction terms, such as the Hubbard and Heisenberg interactions, can be decomposed in terms of the generators listed above and therefore have invariant many-body scar states.

\begin{table}[t!]

	\begin{center}
	\resizebox{\columnwidth}{!}{%
		\begin{tabular}{|c|c|c|c|c|c|c|c|c|}
		 \multicolumn{1}{c}{}& \multicolumn{4}{c}{annihilate $\ket{n^\zeta}$}  &  \multicolumn{3}{c}{}  &  \multicolumn{1}{c}{$\widetilde{\rm SU}'(N)$} \\
		 \multicolumn{1}{c}{}& \multicolumn{4}{c}{$\overbrace{\rule{4cm}{0pt}}$} & \multicolumn{3}{c}{} &  \multicolumn{1}{c}{$\downarrow$}  \\
			\hline
		$\ket{}$ 		&	$T_{ij}$ \eqref{hop:T} &  $K_i$ \eqref{eq:Ki} 	& $\eta^\alpha$ \eqref{hop:eta} &  $T^{\rm O}_{ij}$ \eqref{hop:tilT} 
		 & $\tilde{T}^{A}_{ij}$ \eqref{hop:Ttilalpha} 	& $M_i$  	& $Q_A$ 	& $T'$ \eqref{hop:T'}  \\
		\hline
		$\ket{n^\zeta}$ 	& 0 	& 0 	& 0 	& 0  & - & $*$ & - & 0 \\
		\hline
		$\ket{n^\eta}$ 	& - 	& $*$	& - 	& 0    &0  & 0 &0 & -\\
		\hline
		$\ket{n'^\eta}$      & -	& $*$	& -	& -	& - & 0 & 0 & 0 \\
		\hline
		\multicolumn{4}{c}{}	& \multicolumn{4}{c}{$\underbrace{\rule{4cm}{0pt}}$} & \multicolumn{1}{c}{} \\
		\multicolumn{4}{c}{}	& \multicolumn{4}{c}{annihilate $\ket{n^\eta}$} & \multicolumn{1}{c}{}	
	\end{tabular} 
	}
	\end{center}
	\caption{ Action of generators on the invariant states. The entries $0$ denote the states annihilated by them. The entries $*$: These operators act with a constant when summed over all lattice sites, see Tab. \ref{tab:summedGenOnStates}. "-" means the states are not eigenstates of the corresponding operator. 
	}	
\label{tab:stAnnihilators}
\end{table}

\subsection{Longer-range hopping terms 
\label{sec:SMrealHopAmpl}}

Some classic models, e.g. the Hubbard model, include only the nearest-neighbor hopping term with a real coefficient $t$:
\begin{gather}
H_{nn} = t\sum_{\langle ij \rangle \sigma} (c^\dagger_{i\sigma} c_{j\sigma} + h.c.)\ .
\end{gather}
However, the models that aim to describe realistic materials, such as the high-$T_c$ superconductors, may include the next-to-nearest-neighbor or even longer-range hopping terms with real coefficients. 
It is, therefore, interesting to inquire if the deformation by such additional hopping terms preserves the families of scar states.
In this section we discuss when such terms are generators of the relevant symmetry groups and annihilate the paired states $\ket{n^\eta}'$. 

Any Hermitian spin-independent hopping terms on any lattice (including $T'$, $T^{\rm O}_{ij}$) are special cases of the complex-amplitude hopping $T_{ij}$ \eqref{hop:T}. Therefore, they belong to an SU$(N)$ algebra and annihilate the $\ket{n^\zeta}$ subspace.
The nearest-neighbor hopping with real amplitude \eqref{hop:T'} on a bipartite lattice is a generator of $\widetilde{\rm SU}(N)'$ and annihilates $\ket{n^\eta}'$.

A complication with the longer-range real-amplitude hopping 
\begin{gather}
	T^r_{i j} = \sum_{\sigma} (c^\dagger_{i\sigma} c_{j\sigma} + c^\dagger_{j\sigma} c_{i\sigma})\ , \label{realhop}
\end{gather}
comes from the fact that a commutator of such terms may not always be expressed as $T^r_{kl}$ alone.
Yet, in some special cases, a constraint on the lattice type, hopping range or boundary conditions allows this algebra to be closed.
For example $T'_{\langle i j \rangle}$ is restricted to the nearest-neighbor hopping on a bipartite lattice where we can perform a transformation \eqref{eq:Rtransform} on $c_k$ which maps $t \to i t$. 
Then the hopping terms become generators \eqref{hop:tilT} of SO$(N)$, which is important for the argument that the states $\ket{n^\eta}$ are scars. 

However, if the next-to-nearest-neighbor (nnn) hopping terms are added, the transformation keeps those hopping coefficient real. As a result, the O$(N)$ invariant states $\ket{n^\eta}$ are not scars in presence of the nnn hopping terms.
Indeed, combining the nearest- and next-nearest-neighbor hopping does {\it not} lead to a closed algebra. For example, let us consider the following set of hopping terms connecting sites $1,2,3$ in a linear chain. 
\begin{gather}
	T_1 = \sum_{\sigma} (c^\dagger_{1\sigma} c_{2\sigma} + {\rm h.c.})\ , \quad T_2 = \sum_{\sigma} (c^\dagger_{2\sigma} c_{3\sigma} + {\rm h.c.})\ , \\ \notag 
	T_3 = \sum_{\sigma} (c^\dagger_{1\sigma} c_{3\sigma} + {\rm h.c.}) \ . 
\end{gather}
The hoppings $T_{1}$ and $T_2$ are nearest-neighbor, while $T_3$ is next-nearest-neighbor. Then 
\begin{gather}
[T_1,T_2] = 
\sum_{\sigma} (c^\dagger_{3\sigma} c_{1\sigma} - c^\dagger_{1\sigma} c_{3\sigma})\ .
\end{gather}
It is not of the form of $T_3$; therefore, the subalgebra of hopping terms with real coefficients does not close. 

To avoid this issue, we may use a bipartite lattice and restrict the allowed hopping terms. A closed SO$(N)'$ algebra is formed by the "even" long-range real-amplitude hopping that contains exclusively the terms connecting the sites belonging to different groups of the bipartite lattice (red-to-blue) in Fig. \ref{fig:snake}, meaning that it is a hopping over $2k$ neighbours, where $k$ is a non-negative integer. The number of the nearest neighbors hopped over by the term $t_{ij}  \sum_\sigma c^\dagger_{i\sigma} c_{j\sigma}$ is determined by the lattice as the smallest number of connected sites that need to be visited when travelling from site $i$ to site $j$. The even-nearest-neighbour hopping is shown in red lines in Fig. \ref{fig:snake} and includes the nearest-neighbour hopping as a subset. Such hopping annihilates $\ket{n^\eta}'$ as a generator of SO$(N)'$ and $\ket{n^\zeta}$ as a generator of SU$(N)$.

The algebra SO$(N)'$ can also be constructed on a bipartite lattice when both even- and odd-nearest-neighbour hopping are present in the system. In this case, we must require that all the even-nearest-neighbor hopping terms have real amplitudes while all the odd-nearest-neighbor terms have imaginary amplitudes (for example, in Fig. \ref{fig:snake} we may include the nearest-neighbor hoppings with a real amplitude and the shortest diagonal hoppings with an imaginary amplitude). Such a hopping will again annihilate both $\ket{n^\zeta}$ and $\ket{n^\eta}'$ subspaces. An example of this scenario is the Haldane model \cite{HaldaneModel1988} discussed in Sec. \ref{sec:HHmodel}.

\subsection{Electromagnetic field}

In the absence of an electromagnetic field, the hopping terms are typically taken to have real coefficients. 
The presence of a (possibly time-dependent) electromagnetic field in a lattice model introduces the phase factors $e^{i\alpha_{ij}} c_{i\sigma}^\dagger c_{j\sigma}$. As discussed above, such spin-independent hopping terms with a complex amplitude $T_{ij}$ are generators of SU$(N)$ and therefore annihilate the $\ket{n^\zeta}$ states.

The only effect of an electromagnetic field on the $\ket{n^\zeta}$ states comes from the coupling between the magnetic field and the spin of the electrons:
$\delta H= - \mu \vec Q \cdot \vec B$, which results in a linear in $B$ and equidistant splitting of the $\ket{n^\zeta}$ states according to the projection of their total spin and changes the "revivals" period. 
Other families of the invariant states are mixed by the electromagnetic field with the rest of the spectrum.
Because $\ket{n^\zeta}$ forms the maximum-spin representation of SU$(2)_{spin}$ these states get split the most by the magnetic field w.r.t. to all other states in the Hilbert space. If all other terms in the Hamiltonian are bounded this means that in a strong enough magnetic field an $\ket{n^\zeta}$ state can always be made the ground state.

For some particular configurations (for example a closed 1D chain with a magnetic field of $\pi$ through it), the modified hopping amplitude may be purely imaginary. In such cases, the hopping $T^{\rm O}_{ij}$ \eqref{hop:tilT} is a generator of SO$(N)$, which makes the states $\ket{n^\eta}$ insensitive to the magnetic field (they also have zero spin).

\section{Interaction terms \label{sec:interTerms}}

Let us consider an arbitrary lattice with $N$ sites on which spin-$\frac12$ electrons are placed, and let $i$ and $j$ refer to two sites on this lattice. We would like to rewrite the commonly used electron-electron interaction terms in the form $H_0+OT$ and analyze their action on the three invariant subspaces.

\subsection{Hubbard interaction \label{sec:HubbardInt} }

In the Hubbard model, two electrons interact only when they are located on the same site $i$. 
The Hubbard interaction at site $i$ may be written in terms of the generators $K_i$ of group U$(N)$:
\begin{gather}
\label{eq:hubi}
H^{Hub}_i = n_{i\uparrow} n_{i\downarrow} = \frac{1}{2} \left (n_i^2 - n_i \right ) =  \frac{1}{2} \left (K_i^2 + K_i \right )\ .
\end{gather}
This annihilates the $\ket{n^\zeta}$ states. While neither $\ket{n^\eta}'$ nor $\ket{n^\eta}$ are eigenstates of \eqref{eq:hubi}, they are eigenstates of the Hubbard interaction summed over the full lattice that we write using \eqref{eq:hubi}, \eqref{eq:KiThroughMi} and \eqref{eq:Ki} as
\begin{gather}
\sum_i H^{Hub}_i = \frac{1}{2} \left ( Q -N + \sum_i K_i^2 \right ) = \frac{1}{2} \left (Q - \sum_i M_i^2 \right )\ .
\label{eq:hubiForNO}
\end{gather}
Since the local magnetization $M_i$ annihilates the states $\ket{n^\eta}$ and $\ket{n^\eta}'$, we have 
\begin{gather}
\sum\limits_i H^{Hub}_i \ket{n^\eta}' = n \ket{n^\eta}' = \frac{Q}{2} \ket{n^\eta}', \notag\\
\sum\limits_i H^{Hub}_i \ket{n^\eta} = n \ket{n^\eta}\ ,
\end{gather}
which implies that the energies of these scar states have equal spacing. This is related to the fact that the pseudospin generators $\eta^{\pm}$ do not commute with the Hamiltonian of the Hubbard model \cite{etaPairingYang89,yang1990so,ZhangHubbardSO41991}.

A slightly modified Hubbard interaction
\begin{gather}
\tilde{H}^{Hub}_i = \left(2n_{i\uparrow} - 1\right)\left(2n_{i\downarrow} - 1 \right) =  (-1)^{n_i},
\end{gather}
can be cast in the form
\begin{gather}
\tilde{H}^{Hub}_i = 1- 2 M_i^2 =2K_i^2-1 
\end{gather}
and has the advantage that all three invariant families are eigenstates on {\it every} site:
$\tilde{H}^{Hub}_i \ket{n_\zeta} = -\ket{n_\zeta}$; $\tilde{H}^{Hub}_i \ket{n_\eta}= \ket{n^\eta}$;  $\tilde{H}^{Hub}_i \ket{n_\eta}'= \ket{n^\eta}'$.

\subsection{Density-density interaction \label{sec:densityDensityInter}}

The density-density interaction at different sites, $H^{dd}_{ij} = n_i n_j$, can also be written in terms of the generators (\ref{eq:Ki}):
\begin{gather}
\label{eq:densDens}
H^{dd}_{ij} =  \left(K_i + 1 \right)\left(K_j + 1 \right)\ .
\end{gather}
For the SU$(N)$-invariant states $\ket{n^\zeta}$, we have $H^{dd}_{ij}\ket{n^\zeta} = \ket{n^\zeta}$. Neither $\ket{n^\eta}$ nor $\ket{n^\eta}'$ states are eigenstates of $H^{dd}_{ij}$ since $K_i$ does not act on them with a definite value.

The generalized density-density interaction reads 
\begin{gather}
 H^{gdd}_{ij} = \sum_{\sigma, \sigma'} V^{\sigma \sigma'}_{ij} n_{i\sigma}n_{j\sigma'} = \notag \\ \alpha_{ij} n_i n_j + \beta_{ij} n_i M_j+ \gamma_{ij} M_i n_j + \delta_{ij} M_i M_j,
\label{eq:genDenDenV}
\end{gather}
and $M_i \ket{n^\eta}' = M_i \ket{n^\eta} = 0$ (see Tab \ref{tab:stAnnihilators}).  
$M_i$ does not annihilate $\ket{n^\zeta}$ and mixes them with the non-singlet states. However $\ket{n^\zeta}$ are eigenstates of the total magnetization: $\sum\limits_i M_i\ket{n^\zeta} = (N-2n)\ket{n^\zeta}$.

Similarly, $\ket{n^\eta}$ and $\ket{n^\eta}'$ are not eigenstates of $n_i$, but for the total fermion number $Q$ we have $Q\ket{n^\eta}' = 2n\ket{n^\eta}'$ and $Q\ket{n^\eta} = 2n\ket{n^\eta}$. For $\ket{n^\zeta}$ we have $n_i\ket{n^\zeta} = \ket{n^\zeta}$.
The above observations mean that  $H^{gdd}_{ij}$ annihilates $\ket{n^\eta}$ and $\ket{n^\eta}'$ states when $\alpha_{ij}  =  0$ and leaves $\ket{n^\zeta}$ unchanged when $\beta_{ij}=\gamma_{ij}=\delta_{ij} =0$; for example $\alpha_{ij}n_in_j \ket{n^\zeta} = \alpha_{ij} \ket{n^\zeta}$.

When the coefficients in Eq. \eqref{eq:genDenDenV} are translation-invariant, $V^{\sigma,\sigma'}_{ij} = V^{\sigma,\sigma'}_{i-j}$, the $\ket{n^\zeta}$ become eigenstates of $\beta$ and $\gamma$ terms:
\begin{gather*}
\sum_i \beta_k n_i M_{i+k} \ket{n^\zeta} = \beta_k \sum_i M_{i+k} \ket{n^\zeta} = \beta_k (N-2n) \ket{n^\zeta}.
\end{gather*}
We note that Ref. \cite{moudgalya2020etapairing} demonstrated that for $\sum_{\sigma} V^{\sigma,\sigma'}_{ij} = 0$ (satisfied by the  $\beta,\gamma,\delta$ and violated by $\alpha_{ij}$ terms) \eqref{eq:genDenDenV} respects $\eta$-symmetry and admits the $\ket{n^\eta}'$ states as scars.

\subsection{Heisenberg interaction\label{sec:HeisInt}}

The Heisenberg interaction $H^{Heis}_{ij} = \vec{S}_i \cdot \vec{S}_j$ couples spins on two different sites, $i$ and $j$. 
Using the expression \eqref{eq:SiInFerm} for the spin operator in the Hilbert space of spin-1/2 complex fermions, we find 
\begin{gather}
4\vec{S}_i \cdot \vec{S}_j=\left(c^\dagger_{i \uparrow } c_{i \uparrow} - c^\dagger_{i \downarrow} c_{i \downarrow}\right)\left(c^\dagger_{j \uparrow} c_{j \uparrow} 
- c^\dagger_{j \downarrow} c_{j \downarrow}\right)\notag\\+ \left(c^\dagger_{i \uparrow} c_{i \downarrow} + c^\dagger_{i \downarrow} c_{i \uparrow}\right)\left(c^\dagger_{j \uparrow} c_{j \downarrow} + c^\dagger_{j \downarrow} c_{j \uparrow}\right) \notag\\
-\left(c^\dagger_{i \uparrow} c_{i \downarrow} - c^\dagger_{i \downarrow} c_{i \uparrow}\right)\left(c^\dagger_{j \uparrow} c_{j \downarrow} - c^\dagger_{j \downarrow} c_{j \uparrow}\right) = \notag\\
 \sum_{\alpha,\beta} \bigg (-  c^\dagger_{i\alpha} c_{j\alpha} c^\dagger_{j\beta}  c_{i\beta}-  c^\dagger_{j\alpha} c_{i\alpha} 
c^\dagger_{i\beta}  c_{j\beta}\notag \\
- c^\dagger_{i\alpha} c_{i\alpha} c^\dagger_{j\beta} c_{j\beta}
 +  c^\dagger_{i\alpha} c_{i\alpha}   +  c^\dagger_{j\alpha} c_{j\alpha} \bigg ) \ .
\end{gather}
This may be written as
\begin{gather}
\label{eq:HHeisij}
\vec{S}_i \cdot \vec{S}_j = \frac14 + C_{ij},
\end{gather}
where\\
\begin{gather}
C_{ij}  = - \frac{1}{4} (E_{ij}E_{ji} +  E_{ji}E_{ij}  + K_i K_j)
\label{eq:THeis}
\end{gather}
and we introduced the off-diagonal generators of SU$(N)$:
	\begin{gather}
	E_{ij} =\sum_\alpha c^\dagger_{i\alpha} c_{j\alpha}\ , \qquad i\neq j\ .
\label{eq:defEij}
	\end{gather}
Eq. \eqref{eq:HHeisij} represents the $H_0+OT$ decomposition of the Heisenberg interaction w.r.t. to the group U$(N)$.
It follows that the U$(N)$-invariant states $\ket{n^\zeta}$ are degenerate eigenstates of $\vec{S}_i \cdot \vec{S}_j$ on any lattice, in any dimension and for any $i\neq j$ with energy $\frac14$.
We also note that the Heisenberg interaction 
annihilates the states $\ket{n^\eta}$ and $\ket{n^\eta}'$ because they are annihilated by the local spin operators $S_i^A$ (see Eq. \eqref{eq:localspincomm} and the discussion following it).

\subsection{Symmetry-breaking perturbation \label{sec:symmBreakPert}}

To highlight the ergodicity-breaking properties of the invariant states and to be able to tune to the fully chaotic regime we will consider a simple symmetry-breaking term of the $OT$ form that we write down for a rectangular lattice in two dimensions, where $i$ labels the horizontal and $j$ the vertical direction:
\begin{gather}
H^{p}_{H} = \sum_{i,j} r_{ij} (\tilde{M}_{ij}+ \tilde{M}_{(i+1)j}) S^{\rm hor}_{ij} + \notag\\ q_{ij} (\tilde{M}_{ij}+ \tilde{M}_{i(j+1)}) S^{\rm vert}_{ij}\ ,
\label{eq:Hpert}
\end{gather} 
where $r_{ij}, q_{ij} \in \left[0,1\right]$ are real random numbers and
\begin{gather}
\label{eq:assymMagn}
\tilde{M}_{ij} = r_M 
 c^{\dagger}_{ij\uparrow}c_{ij\uparrow} - q_M
 c^{\dagger}_{ij\downarrow}c_{ij\downarrow}\  ,\\
S^{\rm hor}_{ij} = \sum_\sigma c^{\dagger}_{(i+1)j\sigma}c_{ij\sigma} + {\rm h.c.}\ ,\notag \\ 
S^{\rm vert}_{ij} = \sum_\sigma c^{\dagger}_{i(j+1)\sigma}c_{ij\sigma} + {\rm h.c.} \notag \ ,
\end{gather}
where $r_M, q_M$ are also real random numbers.

Both $S^{\rm hor}_{ij}$ and $S^{\rm vert}_{ij}$ are special cases of the hopping $T'_{\braket{ij}}$ \eqref{hop:T'} which is simultaneously a generator for SU$(N)$ and $\widetilde{\rm SU} (N)' $ groups (see Sec \ref{sec:nOnU} and Tab. \ref{tab:stAnnihilators}). Therefore the full perturbation term $H^{p}_{H}$ is of the pure $OT$ form and annihilates two invariant families: $\ket{n^\zeta}$ and $\ket{n^\eta}'$. It can be added to any model supporting these states as scars without changing their energy but breaking all symmetries except the particle number conservation.

We note that the symmetry-breaking terms such as \eqref{eq:Hpert} also lead to the absence of any dynamical symmetries that could otherwise cause persistent oscillations of local observables \cite{Buca2020DynamicalSymmetriesHeisenbergMagnet}.

\subsection{Non-Hermitian perturbation}

A non-Hermitian Hamiltonian may result under certain approximations from reducing a closed full system to an effective description of an open sub-system. The non-Hermitian Hamiltonian may not conserve the norm of the state which corresponds to the probability leaking out or into the open system. While the derivation of such an effective description is beyond the scope of this work we show here that the invariant states remain stable and decoupled also in non-Hermitian systems as long as the non-Hermitian Hamiltonian has the form $H_0+OT$. We consider the following non-Hermitian $OT$ term
\begin{gather}
\label{eq:nonHermpert}
H^{p}_{nH} = \sum_{i,j} (\tilde{M}_{ij} - \tilde{M}_{(i+1)j}) S^{\rm hor}_{ij} + q_{ij} (\tilde{M}_{ij} - \tilde{M}_{i(j+1)}) S^{\rm vert}_{ij}\ . 
\end{gather} 
This only differs by a 
minus sign from Eq. \eqref{eq:Hpert}, and thus also annihilates the same two invariant families $\ket{n^\zeta}$ and $\ket{n^\eta}'$.
This operator is invariant under complex conjugation, and its eigenvalues form complex conjugate pairs. According to the classification of Ref \cite{UedaPRXNonHermitianClassification,UedaPRRNonHermitianClassification}, the model falls into the Ginibre symmetry class AI, and therefore will have the same level statistics as the Ginibre GUE (GinUE) distribution \cite{ginibre1965statistical}.

Another possibility to obtain a non-Hermitian term of the OT form annihilating invariant states is by taking only a "half" of the hoppings \eqref{hop:T}, \eqref{hop:tilT}, \eqref{newgen}, \eqref{hop:Ttilalpha}, \eqref{hop:T'}, without the hermitian conjugate.
A 2D modified Hubbard model of this form
\begin{gather}
\label{eq:nonHermHubbard}
H^{Hubbard}_{nH} = \sum_{i,j,\sigma} \bigg (t_1 (c^\dagger_{ij\sigma} c_{(i+1)j\sigma}+  c^\dagger_{ij\sigma} c_{i(j+1)\sigma})\nonumber \\
+ t_2 (c^\dagger_{(i+1)j\sigma} c_{ij\sigma}+  c^\dagger_{i(j+1)\sigma} c_{ij\sigma})\bigg) + U\sum_{i,j} n_{ij\uparrow}n_{ij\downarrow} 
 \ ,
\end{gather}
where $t_1\neq t_2$ are real parameters, was recently shown \cite{2021arXiv210606192H} to be amenable to quantum Monte-Carlo simulations.
Since such non-Hermitian hopping terms are certain generators of the SU$(N)$ group, they annihilate the $\ket{n^\zeta}$ states.

 \section{Models with invariant scars}

\begin{table*}

\begin{center}
\resizebox{\textwidth}{!}{%
\begin{tabular}{|c|c|c|c|c|c|c|}
\hline
Model 							& $H_0$							& $H_0^\zeta=OT^{\eta}$							& $O \cdot T$																& $\ket{n^\zeta}$ 	& $\ket{n^\eta}$			& $\ket{n^\eta}'$														 \\
\hline
Hubbard				
			& $\left(\frac{U}{2} -\mu \right) Q  $ 	& 	$-\frac{UN}{2}$		& $ t\sum\limits_{\braket{ij}} T'_{\braket{ij}}  + \frac{U}{2} \sum\limits_i K_i^2  $    						&  $\checkmark$ 	&  & $\checkmark$	
								  	\\
\hline
Heisenberg 						& 			&	$  \frac{J}{4} N^{nn}_1  $				& $ J\sum\limits_{\braket{ij}} C_{ij} $										& $\checkmark$ 	&  $\checkmark$ 	& $\checkmark$ 			   	 \\

\hline
$J_1-J_2$							& 	&	$   \frac{ J_1 N^{nn}_1 + J_2 N^{nn}_2 }{4}  $ & $  J_1 \sum\limits_{\braket{ij}} C_{ij} + J_2 \sum\limits_{\braket{\braket{kl}}} C_{kl}  $ &$\checkmark$ 	&  $\checkmark$	& $\checkmark$	 \\

\hline
Haldane-Shastry 					&			&	$ \frac{\pi^2(N^2-1)N}{24 N^2} $			& $ - \frac{\pi^2}{N^2} \sum\limits_{n<n'}  \frac{ C_{nn'} }{ \sin^2{ \left(\frac{\pi(n-n')}{N}\right)}} $ & $\checkmark$ & $\checkmark$ & $\checkmark$ 		  	 \\

\hline
$t-J$								& 				&	$ \frac{J}{4} N^{nn}_1 $			&$ J \sum\limits_{\braket{ij}}C_{ij}  +  t\sum\limits_{\braket{ij}} T'_{\braket{ij}}$  									& $\checkmark$ 	&  						& $\checkmark$		 \\

\hline
$tJU$							& $ \left(\frac{U}{2} -\mu \right) Q  $	 &  $\frac{J}{4} N^{nn}_1 -\frac{UN}{2}$ 	& $ J \sum\limits_{\braket{ij}}C_{ij} +  t\sum\limits_{\braket{ij}} T'_{\braket{ij}}  + \frac{U}{2} \sum\limits_i K_i^2$ & $\checkmark$ 	&  						& $\checkmark$		 \\

\hline
Hirsch (reduced) \cite{Hirsch1989Reduced}
&  $ \left(\frac{U}{2} -\mu \right) Q $		&	$ -\frac{UN}{2} $	&    $\sum\limits_{\braket{ij}} O^{HR}_{ij} T'_{\braket{ ij}} + \frac{U}{2}\sum\limits_i K_i^2$		&  $\checkmark$ &  &  $\checkmark$		 	 \\

\hline
Hirsch (full) \cite{HIRSCHModelOriginal}	
&  $\frac{U}{2}(Q-N) + V N^{nn}_1$  							&		& $ \sum\limits_{\langle ij \rangle} O^{HF}_{ij}  T'_{\langle ij \rangle}   +   \frac{U}{2}\sum\limits_j K_j^2  + V \sum\limits_{\langle ij \rangle} (K_i K_j + K_i + K_j)  $					 										& $\checkmark$ &  & 		 \\

\hline
Haldane-Hubbard				& $\left(\frac{U}{2} -\mu \right) Q$			& $-\frac{UN}{2} $	& $t_1  \sum\limits_{\braket{ij}} T'_{\braket{ij}}  +  \frac{U}{2} \sum\limits_i K_i^2 + t_2 \sum\limits_{\braket{\braket{kl}} ,\sigma} \left( e^{i\Phi_{kl}} c^\dagger_{k\sigma} c_{l\sigma} + {\rm h.c.} \right) $				&  $\checkmark$ &  $\checkmark^*$ & $\checkmark^{**}$ 	\\
\hline
\end{tabular}
}
\end{center}
\caption{\label{tab:knownModelsAsOT} Well-known models written as $H_0+OT$. The decomposition is different for paired $\ket{n^\eta}$/$\ket{n^\eta}'$ and $\ket{n^\zeta}$ states. The term in the third ($H_0^\zeta=OT^\eta$) column is part of $H_0$ for the decomposition with respect to the U$(N)$ group ($\ket{n^\zeta}$ scars) and a part of $OT$ with respect to $\widetilde{\rm U}(N)$ or $\widetilde{\rm U}'(N)$ groups ($\ket{n^\eta}$ or $\ket{n^\eta}'$ scars). Eqs. \eqref{eq:KiThroughMi} and \eqref{eq:HHeisij} allow to easily switch between the expressions for $\ket{n^\zeta}$ and paired states. All models are considered in the unrestricted Hilbert space that allows any on-site occupations incl. double-occupations. All decompositions are valid in any dimension and on any lattice except for the models that involve the $\ket{n^\eta}'$ states - they only become scars when the lattice is bipartite. $N^{nn}_1$ and $N^{nn}_2$ are the numbers of nearest- and next-nearest-neighbour pairs in a particular lattice. $C_{ij}$ \eqref{eq:THeis} are generators of SU$(N)$. The last three columns indicate which families of the invariant states are many-body states in the corresponding model.\\ *: only when $\Phi=\pi/2$, $t_1=0$. **: only when $\Phi=\pi/2$}
\end{table*}

The interaction terms considered above, as well as their linear combinations, are of the form $H_0+OT$. Together with the generators from Sec \ref{sec:nOnU} they can be used as building blocks for designing Hamiltonians in which some of the invariant states $\ket{n^\zeta}$, $\ket{n^\eta}$, and $\ket{n^\eta}'$ are many-body scars.

Many commonly used models, such as the Hubbard, Heisenberg, and $tJU$ models fall into this class. In Table \ref{tab:knownModelsAsOT} we explicitly re-write them as $H_0+OT$ and indicate the invariant states comprising the scar subspace, the derivations and some more details are given in Appendix \ref{sec:SMDeriveOT}. The presence of $\ket{n^\eta}'$ scars in the extended Hubbard and Hirsch models, that has been demonstrated in the literature \cite{moudgalya2020etapairing,2020MarkMotrEtaPairHub}, is a direct consequence of the group theoretic structure presented in this work. 

Because several families of scars with different symmetries may be present in the same model simultaneously the $H_0+OT$ decomposition is different between $\ket{n^\zeta}$ and the paired $\ket{n^\eta}/\ket{n^\eta}'$ states. However, the difference is only in that a certain constant term (shown in third column of Table \ref{tab:knownModelsAsOT}) is either a part of $H_0$ or $OT$ depending on which group we consider (full separate expressions with respect to each group are given in Appendix \ref{sec:SMDeriveOT}).

To facilitate the design of custom models not listed here we also added Table \ref{tab:summedGenOnStates} to the Appendix which gives the action of other common Hamiltonian terms on the invariant states.

Note that in models that conserve spin, the $\ket{n^\zeta}$ states that comprise a maximum spin-$N/2$ representation of SU$(2)_{spin}$ will not be true scars. To make them such, adding a perturbation breaking spin-conservation (such as \eqref{eq:Hpert} or \eqref{eq:nonHermpert}) is required. Analogously, the states $\ket{n^\eta}'$ only become true scars upon addition of an $\eta$-pairing symmetry breaking perturbation of the $OT$ form.

\subsection{Engineering the location of scars in the spectrum \label{sec:initToScar}}

\subsubsection{Making a scar the ground state \label{Sec:engineer1}}

The energies of the scars and the basis in the invariant subspace are determined by the $H_0$ part of the Hamiltonian, and we can change their position in the spectrum by adding a term to $H_0$ that commutes with it (is diagonal in the basis selected by $H_0$).

The $\ket{n^\zeta}$ states form the maximum spin representation of SU$(2)_{spin}$ and have a definite total spin and its axis projection quantum numbers. This means that a sufficiently strong magnetic field will make the $\ket{n^\zeta}$ state with the largest axis projection the ground state. It also controls the splitting between the $\ket{n^\zeta}$ states and the revivals period. For the basis in Eq. \eqref{stateNu} one would use the $B_z=Q_3$ \eqref{hop:zeta} magnetic field while for the basis \eqref{stateNuTJU} used later in our numerical example we will use the $B_y=Q_2$ field \eqref{hop:zeta}.

The states $\ket{n^\eta}$ and $\ket{n^\eta}'$ have a definite particle number and include the states with maximal and minimal possible particle number (all filled $\ket{N^\eta}$ or $\ket{N^\eta}'$ and all empty $\ket{0^\eta}$ or $\ket{0^\eta}'$). This guarantees that a sufficiently large chemical potential term can be used to make one of these two states the ground state.

\subsubsection{Coupling to the rest of the scar subspace}

Suppose we are at zero temperature and are in the ground state which per the above results is a scar state. Now we turn on for a limited time the ``raising operator" of the corresponding scar family.

In case the ground state is an all-filled  $\ket{N^\eta}$ or $\ket{N^\eta}'$ state we add a term $\eta^- + \eta^+$ to the Hamiltonian (see \eqref{eq:etaPairState} and \eqref{hop:eta}). This may potentially be made by placing the system into direct contact with a superconductor. As a result, the system will be initialized to a state that is a linear combination of $\ket{n^\eta}$ or $\ket{n^\eta}'$. Moreover, since $\ket{n^\eta}$ form an irreducible representation of SU$(2)_\eta$ we can by the action of a group $e^{i t \left(\alpha \eta^- + \alpha^* \eta^+\right)}$ send any initial state, say $\ket{0^\eta}$ or $\ket{N^\eta}$, to any desired linear combination of the paired states. 

In case the ground state is a state with maximal spin $\ket{N^\zeta}$ we add a term $\zeta + \zeta^\dagger$ \eqref{eq:zeta} to the Hamiltonian. This may be implemented by introducing an external magnetic field ($B_x=Q_1$ for basis \eqref{stateNu} and  $B_z=Q_3$ for basis \eqref{stateNuTJU}). As a result the system will be initialized to a state within the $\ket{n^\zeta}$ subspace. The same algebraic argument as for $\ket{n^\eta}$ above shows that with an appropriate choice of magnetic field and interaction time one can turn any initial state in the $\ket{n^\zeta}$ subspace into an arbitrary desired linear combination of $\ket{n^\zeta}$ states.

Finally, we can mix the states from the families $\ket{n^\eta}$ and $\ket{n^\zeta}$ by the simultaneous action of magnetic field and $\eta$ term.

After the above steps the system is initialized to a state in the singlet scar subspace.
 At this point one may choose to lower or turn off the magnetic field or the chemical potential that was used in the first step \ref{Sec:engineer1} to make one of the scar states the ground state.

 It would be interesting to analyze the relation of this scheme to a number of protocols for preparing a paired $\ket{n^\eta}'$ state that rely on dissipation or periodical driving \cite{Zoller2008natPhys,Jaksch2017PRB,Buca2019DrivingInducedEtaPairing,Werner2020PRB}.  
 
 \subsubsection{Low-energy subspace composed of scars only}
 
 We can also arrange for all the low-energy subspace be composed solely of many-body scars. To do that we need to add a non-negative definite operator $P$ to the Hamiltonian that annihilates the scars (and the scars only). Increasing the magnitude of such a term will leave the scars untouched but will push the rest of the spectrum up in energy.

 For the scar states invariant under a particular group we have various options for designing the desired operator $P$.
 For example, $P$ can be chosen to be the quadratic Casimir operator of the corresponding group, as was done in \cite{Pakrouski:2018jcc}.
However, from the point of view of lattice models this is a complicated and non-local operator. A simple choice of a local operator is $P=\sum_{k} (T_k T^\dagger_k)^{l}$, where $T_k$ are the simple root generators of the group \eqref{eq:simroot} and $l>0$ is an integer.

The Hubbard interaction already includes \eqref{eq:hubiForNO} a similar term $P=\sum_i K^2_i$ with $K_i$ the generator of U$(N)$ \eqref{eq:Ki}. It does annihilate the $\ket{n^\zeta}$ scars and pushes non-scar states high in energy for large $U$ at half-filling (where the $H_0$ energy contribution of the Hubbard interaction is zero). However it also ``accidentally" annihilates some of the non-scar states and thus doesn't alone allow to create a scar-only low-energy subspace.

\begin{figure*}
	\centering
	\includegraphics[width=0.49\textwidth]{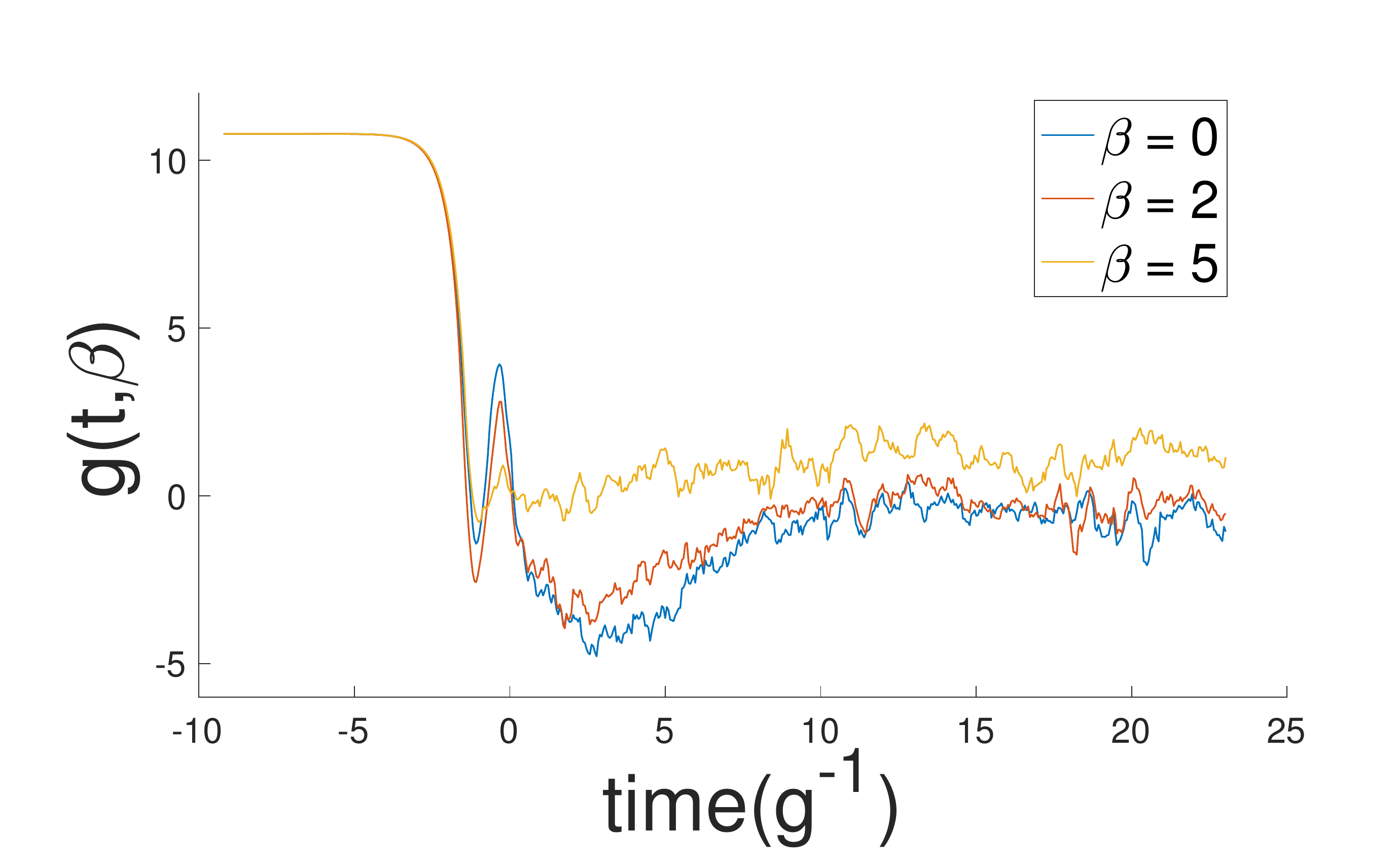}
	\includegraphics[width=0.49\textwidth]{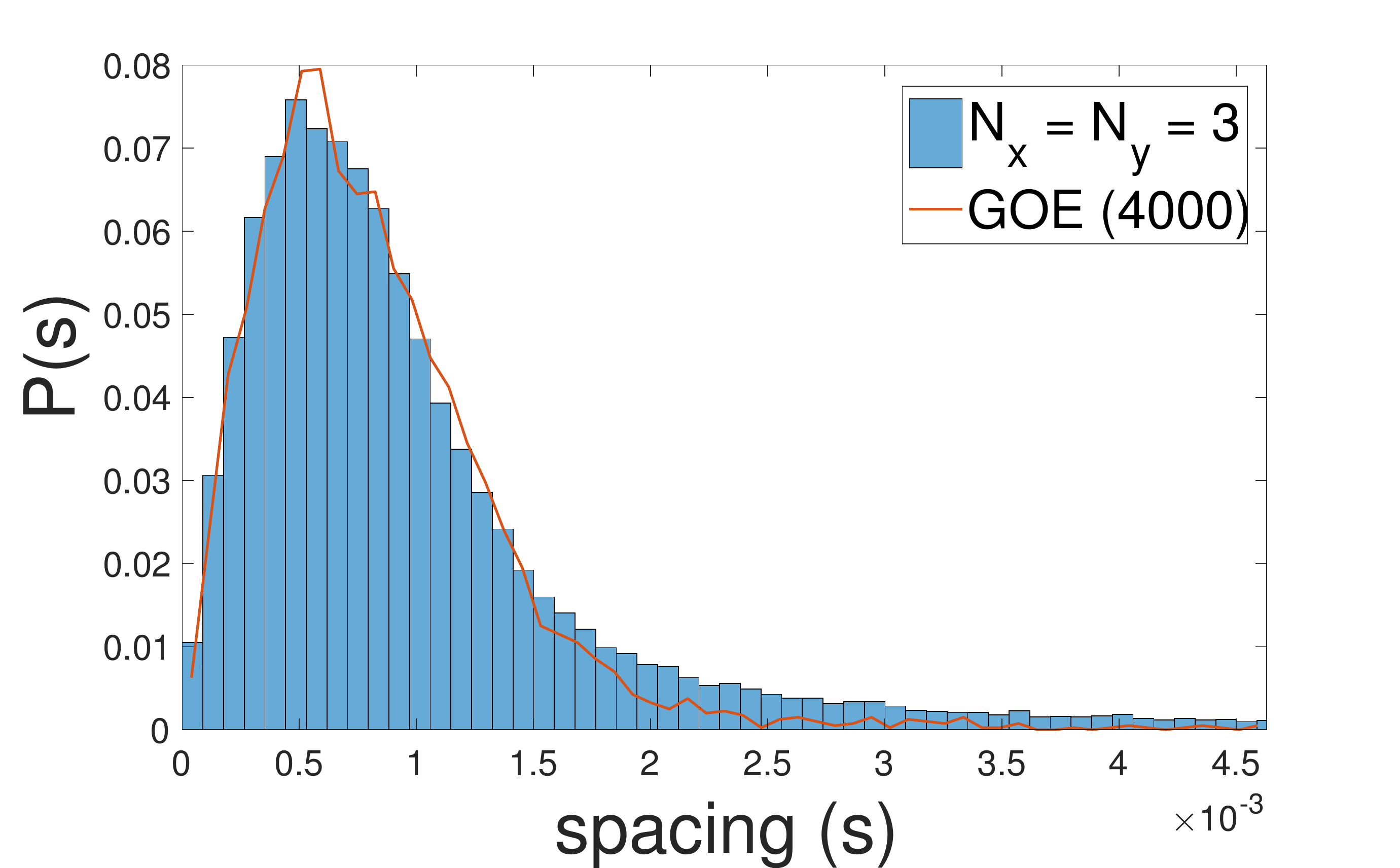}
	\caption{Quantum chaos for Hermitian $tJU$ \eqref{eq:HnumericsHerm} and high magnetic field $\gamma=3.6$. Left: Spectral Form Factor $g(t,\beta)$ for three different temperatures. Right: Level Spacings Distribution $P(s)$. The red curve is the data for the random GOE matrix with size $4000 \times 4000$. This model has $\langle r \rangle = 0.5306$. $\langle r \rangle_{GUE} = 0.6027$ and $\langle r \rangle_{GOE} = 0.5359$ \cite{Atas_2013}. }
		\label{fig:tju_hermSFFandPsG36}
\end{figure*}

\begin{figure*}
	\centering
	\includegraphics[width=0.49\textwidth]{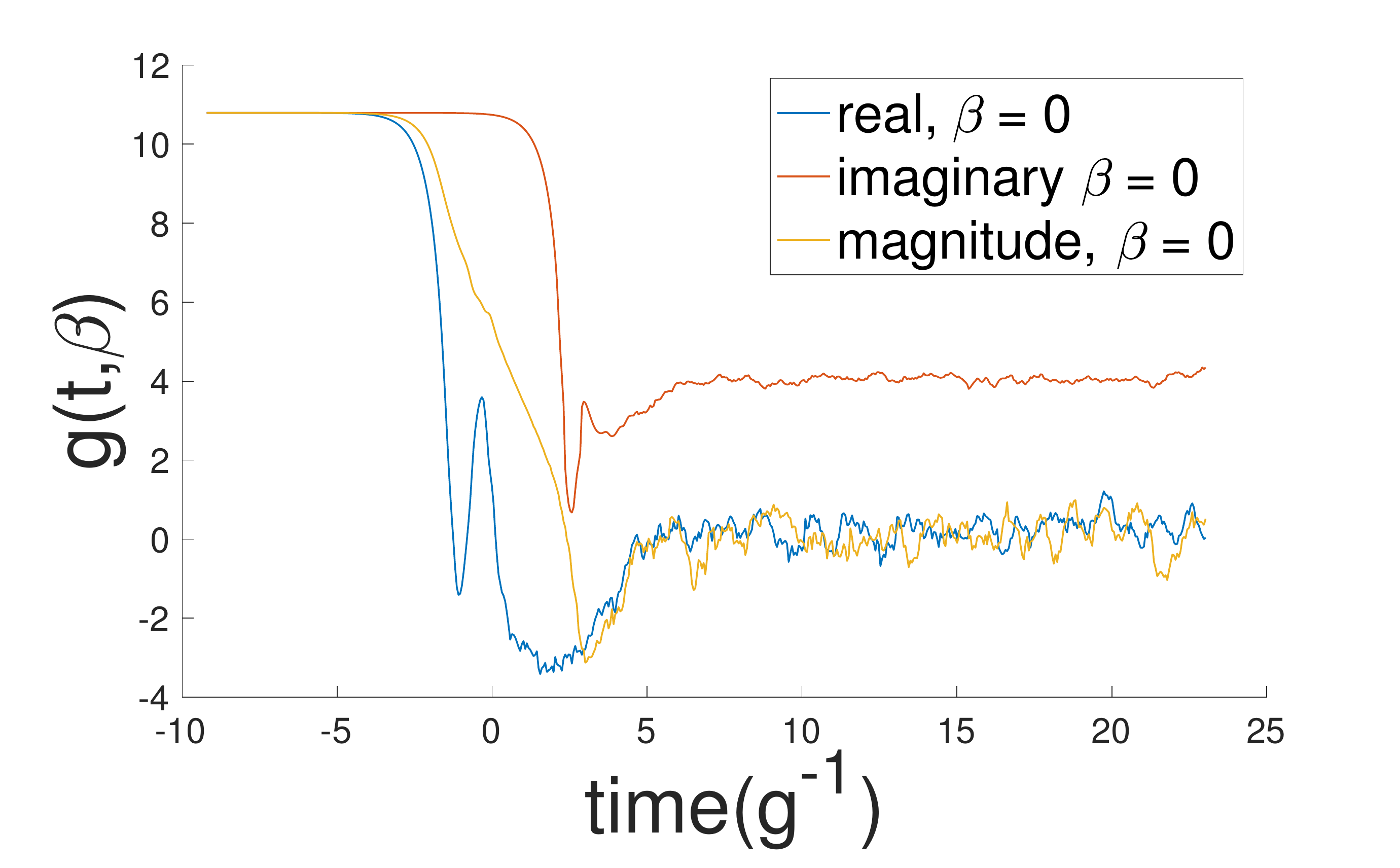}
	\includegraphics[width=0.49\textwidth]{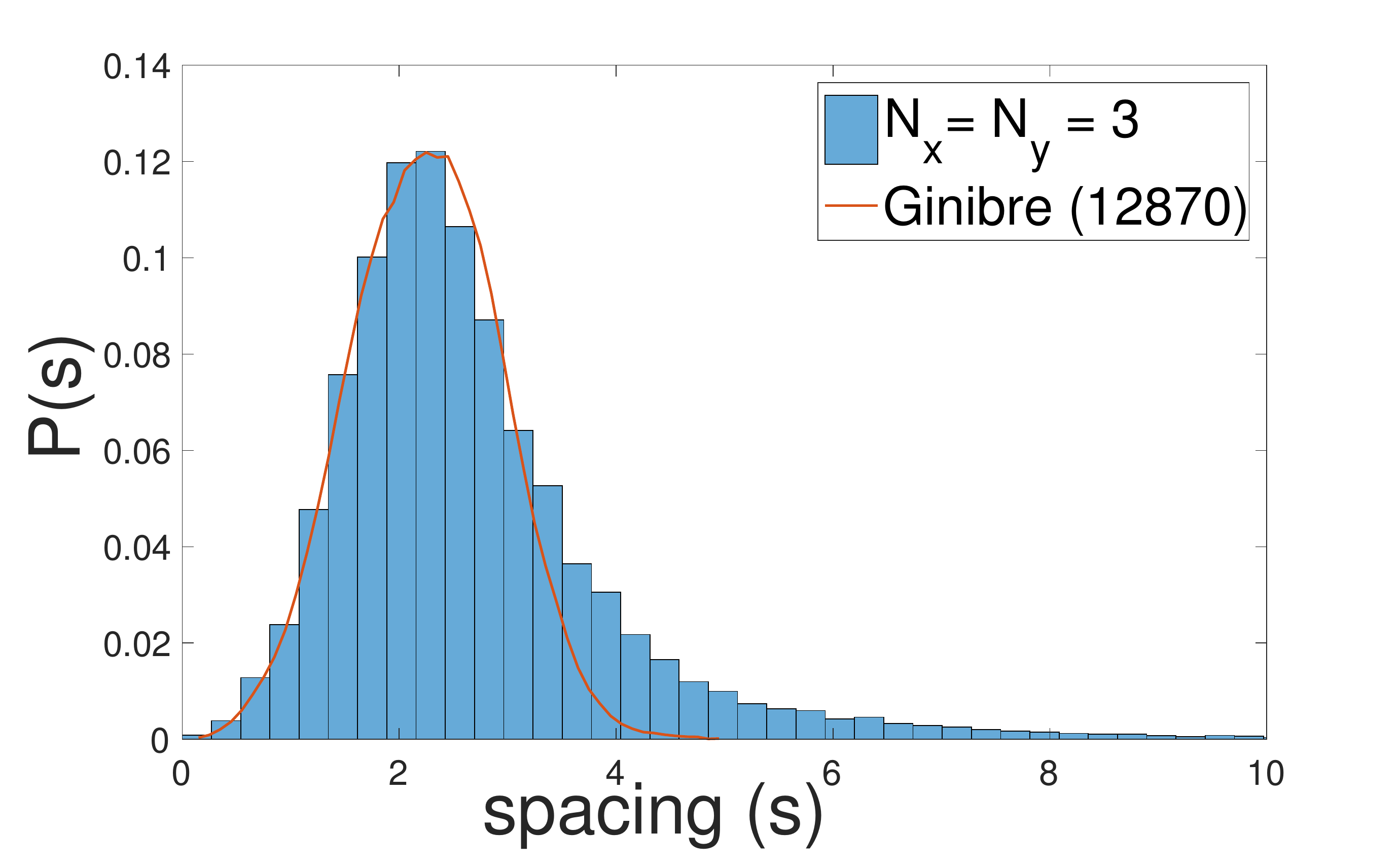}
	\caption{ Quantum chaos for non-Hermitian $tJU$ \eqref{eq:HnumericsNonHerm} and high magnetic field $\gamma=3.6$.
	Left: Spectral Form Factor $g(t,\beta)$ at infinite temperature ($\beta=0$) calculated using only real/imaginary parts of the eigenvalues or their magnitude; Right:  Level Spacing Distribution, $P(s)$. This model has $\langle r \rangle = 0.7378$. For reference, the Ginibre value $\langle r \rangle_{GinUE} \approx 0.74$ was supported numerically in \cite{ProsenPRXNonHermitianR}.}
		\label{fig:tju_nonhermPs}
\end{figure*}

\section{Two-dimensional $tJU$ model }
\label{sec:tJUNumerics}

We illustrate the concepts discussed above using the example of a perturbed $tJU$ model on a $2D$ rectangular bipartite lattice shown in Fig. \ref{fig:snake}. 
The Hamiltonian of the standard $tJU$ model \cite{White2000FirstTJU,Basu2001FirstTJU,Zhang2003FirstTJU} combines the Hubbard and Heisenberg interactions 
\begin{gather}
H^{tJU} = \sum_{\langle ij \rangle \sigma} (t c^\dagger_{i\sigma} c_{j\sigma} + h.c.)  +
 J \sum_{\langle ij \rangle} \vec{S}_i \cdot \vec{S}_j +  \notag\\+ U \sum\limits_i
  n_{i\uparrow}n_{i\downarrow} 
- \mu Q
\label{eq:HtJUstd}
\end{gather}
and can be viewed as a generalization of Hubbard or $t-J$ models relevant for high-$T_c$ superconductivity \cite{MicnasReviewTJULike}. It is typically assumed that $t$ is real and negative, so that the kinetic energy is minimized at zero momentum. 
Certain types of longer-range hopping could be considered in addition without changing the structure of the scar subspace as discussed in Sec. \ref{sec:SMrealHopAmpl}.

The $H_0 + OT$ decomposition of the model 
(\ref{eq:HtJUstd}) with real $t$
can be performed w.r.t. to two groups U$(N)$ and $\widetilde{\rm U} (N)'$ which leads to two families of group-invariant scars $\ket{n^\zeta}$ and $\ket{n^\eta}'$.
In case of U$(N)$ we have 
\begin{gather}
H^{tJU} = \frac{J}{4}N^{nn}_1 +Q\left(\frac{U}{2} -\mu \right)  -\frac{UN}{2} 
+  \notag\\
 t \sum_{\langle ij \rangle} T'_{\langle i j \rangle}  + J \sum\limits_{\braket{ij}} C_{ij}+ \frac{U}{2} \sum\limits_i K^2_i\ ,
\label{eq:HtJUstdNUOT}
\end{gather}
while for the group $\widetilde{\rm U} (N)' $ it is
\begin{gather}
H^{tJU} = Q \left(\frac{U}{2} - \mu\right)+ \notag\\ 
t \sum_{\langle ij \rangle} T'_{\langle i j \rangle}  + J \sum_{\langle ij \rangle} \vec{S}_i \cdot \vec{S}_j  - \frac{U}{2}   \sum_i M_i^2\ ,
\label{eq:HtJUstdNOpOT}
\end{gather}
where we have used the $H_0+OT$ decomposition of individual terms \eqref{hop:T'}, \eqref{eq:hubiForNO}, \eqref{eq:HHeisij} derived earlier. In both equations the first line is $H_0$ and acts on the invariant states with a constant while the second $OT$ line only consists of terms proportional to group generators that annihilate the invariant states. Note that the two expressions \eqref{eq:HtJUstdNUOT} and \eqref{eq:HtJUstdNOpOT} are only different by assigning a certain constant term to $H_0$ or $OT$ as indicated in the decomposition given in Table \ref{tab:knownModelsAsOT}.

We recall that the states invariant under group $G$ become scars in a model that can be written as $H_0+\sum_k O_k T_k$, where $T_k$ are generators of $G$. Note that a generator of any subgroup of $G$ is also a generator of $G$ and can also appear as $T_k$ in the decomposition. This is actually the case in \eqref{eq:HtJUstdNOpOT} (and all other decompositions w.r.t. paired states) where the hopping terms are generator of $\widetilde{\rm SU} (N)'$, while the Heisenberg and magnetization terms involve the local spin $\vec S_i$ that is also a 
part of the full symmetry group of $\ket{n^\eta}'$ (see Sec. \ref{sec:AppSemi}).

\begin{figure}
	\centering

	\includegraphics[width=\columnwidth]{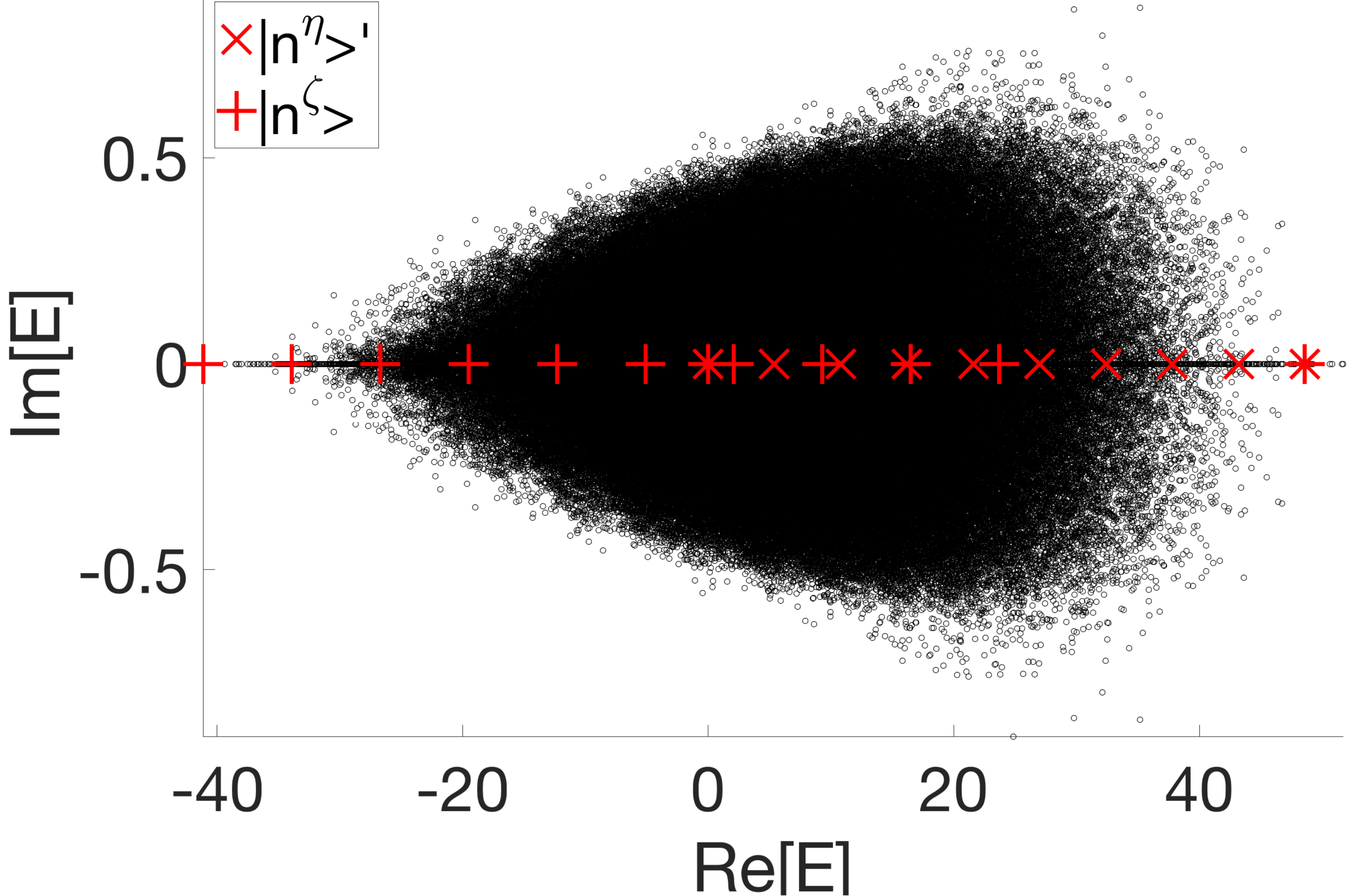}
	
	\caption{Spectrum of the non-Hermitian $tJU$ model \eqref{eq:HnumericsNonHerm}. All the particle number sectors are plotted together. The invariant states shown in red remain at purely real energies.}
		\label{fig:NHSpec}
\end{figure}

The standard $tJU$ model \eqref{eq:HtJUstd} conserves the total physical spin. Therefore the states $\ket{n^\zeta}$ form a separate symmetry sector of this model. To make them true scars we break the total spin conservation by adding a perturbation \eqref{eq:Hpert}. The full {\it Hermitian} Hamiltonian we study numerically reads

\begin{gather}
H^{tJU}_{h} = H^{tJU} + \beta H^{p}_{h}+ \gamma Q_2\ ,
\label{eq:HnumericsHerm}
\end{gather}
where we added a term proportional to the SU$(2)_{spin}$ generator $Q_2$ \eqref{hop:zeta}. It acts as $H_0$ on $\ket{n^\zeta}$ and splits them according to the index $n$: $Q_2\ket{n^\zeta} = (2n-N)\ket{n^\zeta}$). For the SU$(2)_{spin}$-invariant states $\ket{n^\eta}'$ it is of the pure $OT$ form and annihilates them exactly.

Note that by increasing $\gamma$ we can make the scar state with maximum $Q_2$, $\ket{S_1}$ \eqref{stateNuTJU}, the ground state. This may be used to initialize the system to a state from many-body scar subspace as described in Sec. \ref{sec:initToScar}.
The full {\it non-Hermitian} Hamiltonian we consider is
\begin{gather}
H^{tJU}_{nh} = H^{tJU}_{h}  +  \beta_1 H^{p}_{nh},
\label{eq:HnumericsNonHerm}
\end{gather}
where for numerical investigations we set $\beta_1=0.4\beta$.
In both the Hermitian and the non-Hermitian cases, the part of the full Hamiltonian that acts with a constant on the invariant states is 
\begin{gather}
\label{eq:H0tjuNumericsZeta}
H^{ \ket{n^\zeta} }_0 = \frac{J}{4} N^{nn}_1   + \left( \frac{U}{2} -\mu  \right) Q - \frac{UN}{2} 
+ \gamma Q_2 \\
H^{ \ket{n^\eta}' }_0 =\left(\frac{U}{2} -\mu \right) Q.
\label{eq:H0tjuNumericsEta}
\end{gather} 
The states $\ket{n^\eta}'$ are already eigenstates of \eqref{eq:H0tjuNumericsEta} as written in Eq. \eqref{eq:etaPairState} while the states in Eq. \eqref{stateNu} are not the eigenstates of \eqref{eq:H0tjuNumericsZeta}. Instead, the basis in the SU$(N)$-invariant subspace determined by the Hamiltonian \eqref{eq:H0tjuNumericsZeta} reads

\begin{gather}
\label{stateNuTJU}
\ket{n^{\tilde \zeta}_{tJU}} = \frac{\tilde \zeta^n}{2^n \sqrt{\frac{N! n!}{(N-n)!}}} \ket{S_1}\ ,  \\
\ket{S_1}  = \prod_a \frac{c^\dagger_{a1} + i c^\dagger_{a2}}{\sqrt{2}} \ket{0}\ , \nonumber
\end{gather}
where $\tilde \zeta = Q_3 - i Q_1$. 

The energies of the invariant states are given by
\begin{gather}
\label{eq:nonuEnInNum}
E_\eta^{n} = \left(U-2\mu\right) n, \notag\\
E_{\tilde{\zeta}}^{n} =  \frac{J}{4} N_1^{nn} - \mu N + \gamma(2n - N) ,
\label{eq:nonuEnInNum}
\end{gather}
where $n$ - is the index of a state in its respective family \eqref{eq:etaPairState} or \eqref{stateNuTJU}.

The energy of the product state $\ket{0^{\tilde \zeta}_{tJU}}=\ket{S_1}$ in large systems is proportional to $N$ and therefore has good chances to be the ground state at half-filling for $(J/2-\mu-\gamma)<0$. An $\ket{n^\eta}'$ state is more likely to be the ground state when $(U-2\mu)<0$.

Because both Hermitian and non-Hermitian models we consider have an exact $H_0+OT$ decomposition, they have the two families $\ket{n^\zeta}$ and $\ket{n^\eta}'$ as scars for any choices of the coupling constants. The two scar families form two equidistant towers of states with the energies given in \eqref{eq:nonuEnInNum}. Revivals within each individual subspace can be observed for any values of the couplings. However, to see the revivals of an initial state that is a mix of $\ket{n^\zeta}$ and $\ket{n^\eta}'$ subspaces all the gaps between them must have a common divisor which represents a constraint on the choice of the constants $\mu$, $U$, and $\gamma$.

\begin{figure}
	\centering
	\includegraphics[width=\columnwidth]{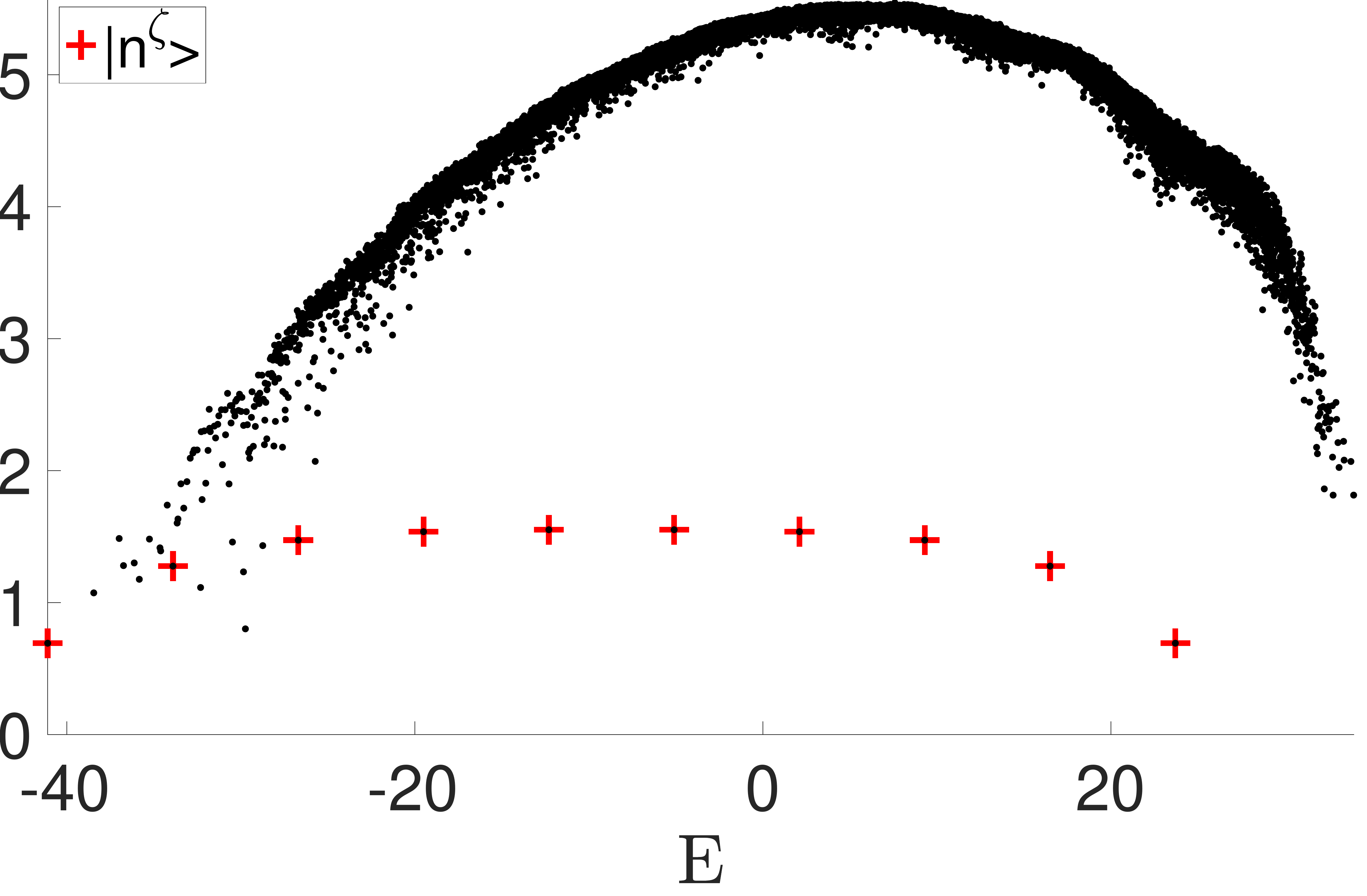}\\
	\includegraphics[width=\columnwidth]{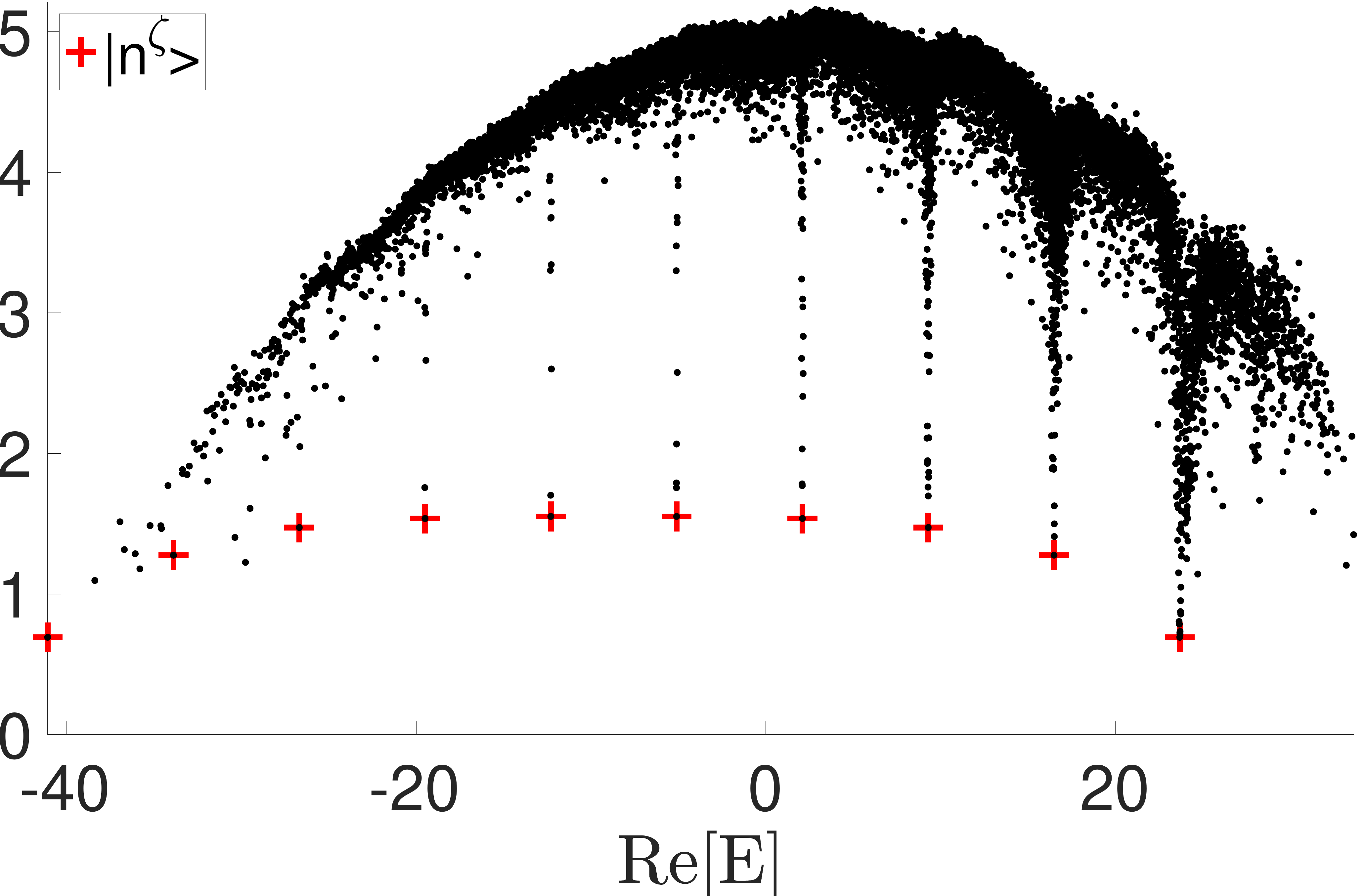}
	\caption{Entanglement entropy computed numerically for the 3$\times$3 system in the half-filling sector $Q=9$. The cut separates the five sites in the first and second rows from the four sites in the second and the third row. 
	Top:
	Hermitian Hamiltonian \eqref{eq:HnumericsHerm} 
	Bottom:
	non-Hermitian Hamiltonian \eqref{eq:HnumericsNonHerm}}
		\label{fig:entropies}
\end{figure}

\subsection{Numerical results}

For the numerical experiment we use the rectangular 3$\times$3 lattice and choose open boundary conditions to make sure the real-amplitude nearest-neighbor hopping is a generator of $\widetilde{\rm U}(N)'$ and annihilates the $\ket{n^\eta}'$ states (boundary condition wouldn't matter in a system with even number of sites). We set $t=-0.4$, $J=1$, $U=8$, $\mu = 1.3$ and $r_M=1.426974$ and $q_M=2.890703$ (see Eq. \eqref{eq:assymMagn}). For $\gamma=1$ this corresponds to the g.s. filling of $\nu=\frac{Q}{2N}=0.44(4)$, 11\% below the half-filling which corresponds to the potentially high-$T_c$-relevant regime \cite{DagottoRMPtjuParams,Abram2017tjuParams}. For our simulation we instead choose $\gamma=3.6$. At this value the half-filled $\ket{S_1}$ state \eqref{stateNuTJU} becomes the ground state. 
This simplifies the initialization of the system to the scar subspace in experiment, which is discussed in detail in Sec. \ref{sec:initToScar}. 
Because $\ket{S_1}$ is a product state, it can alternatively be created by application of a simple gate circuit on each site (see SM of Ref \cite{pakrouski2020GroupInvariantScars}).

\begin{figure*}
	\centering
	\includegraphics[width=0.49\textwidth]{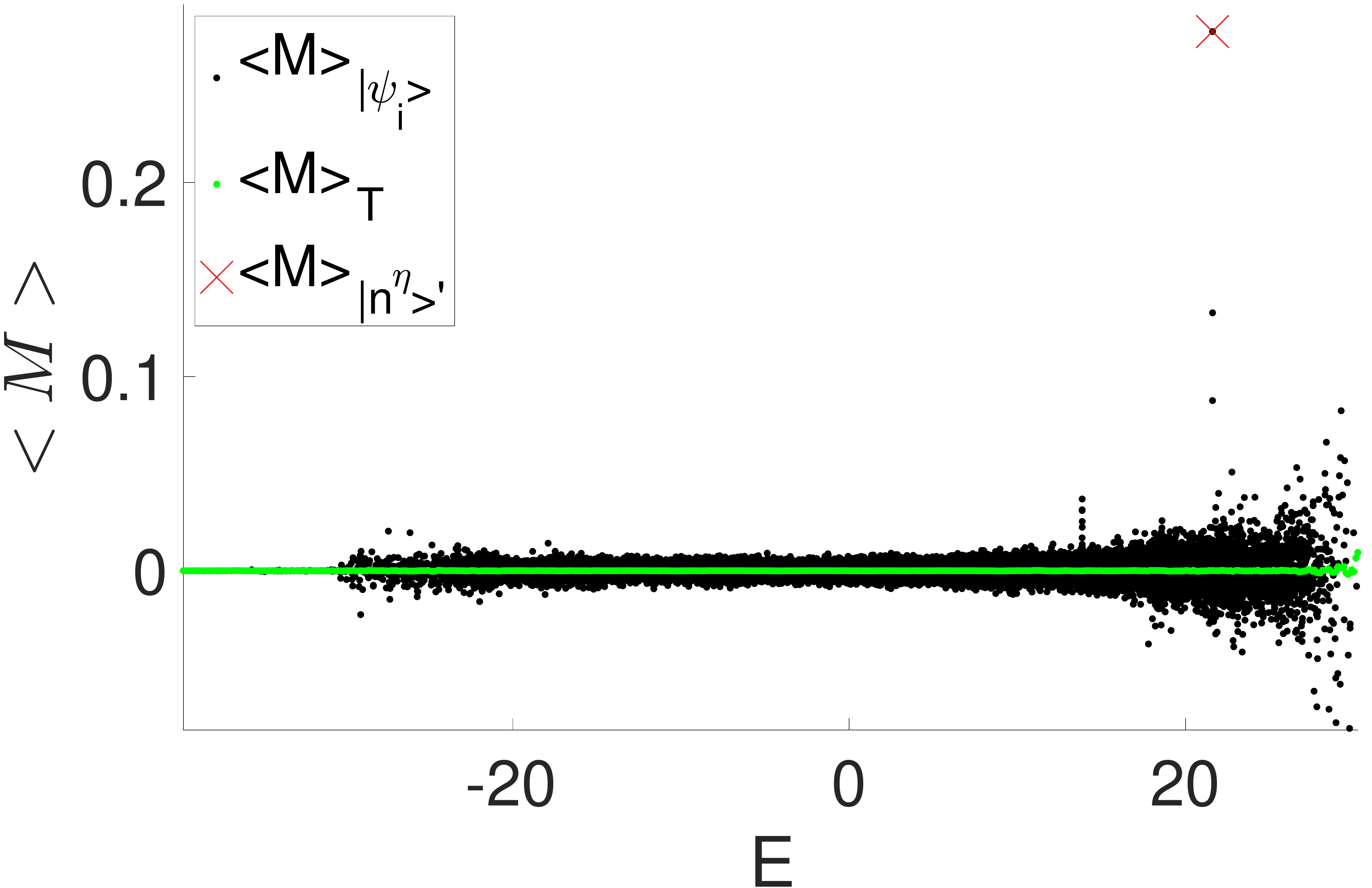}
	\includegraphics[width=0.49\textwidth]{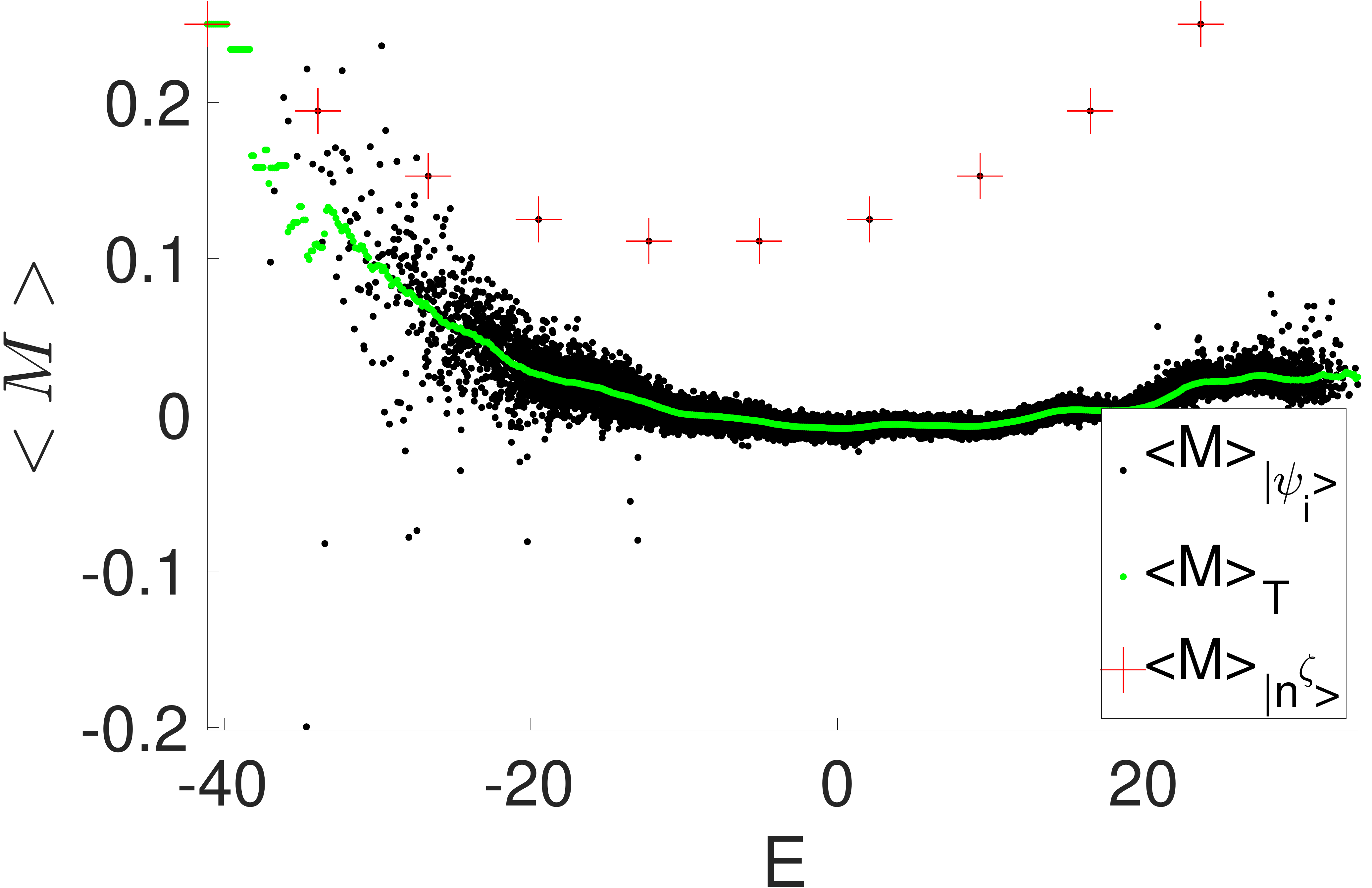}
	\caption{ETH violation by the scar states for the Hermitian Hamiltonian \eqref{eq:HnumericsHerm}. Shown is the expectation value $\braket{\psi_i|M|\psi_i}$ evaluated for every eigenstate $\psi_i$. The green line is the micro-canonical (window) average. Left Panel: particle number sector Q=8, operator $M=G_O$ \eqref{eq:Go}. Right Panel: particle number sector Q=9 (half-filling) $M = G_U$ \eqref{eq:Gu}.}
		\label{fig:ODLROHerm}
\end{figure*}

\begin{figure*}
	\centering
	\includegraphics[width=0.49\textwidth]{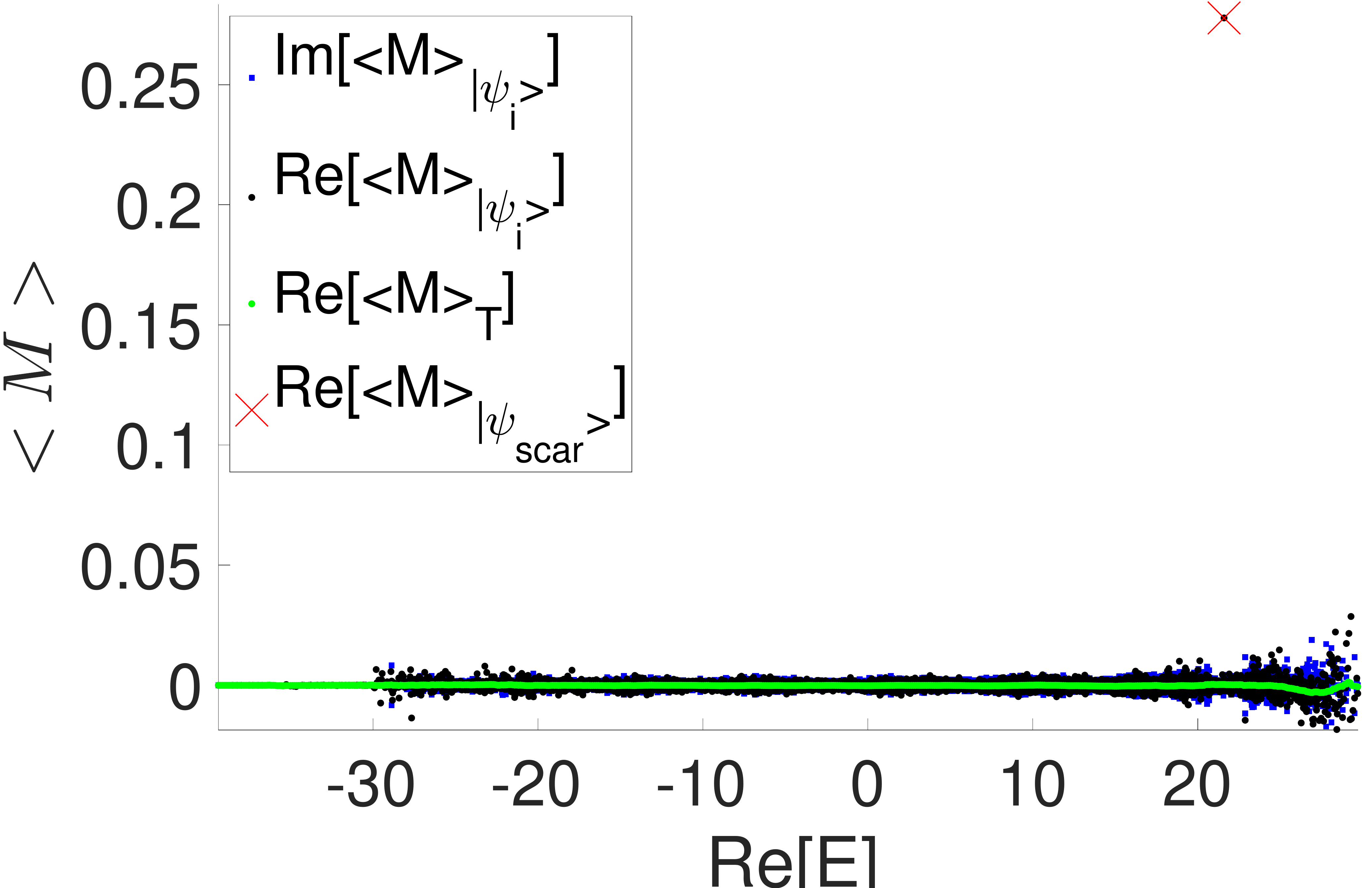}
	\includegraphics[width=0.49\textwidth]{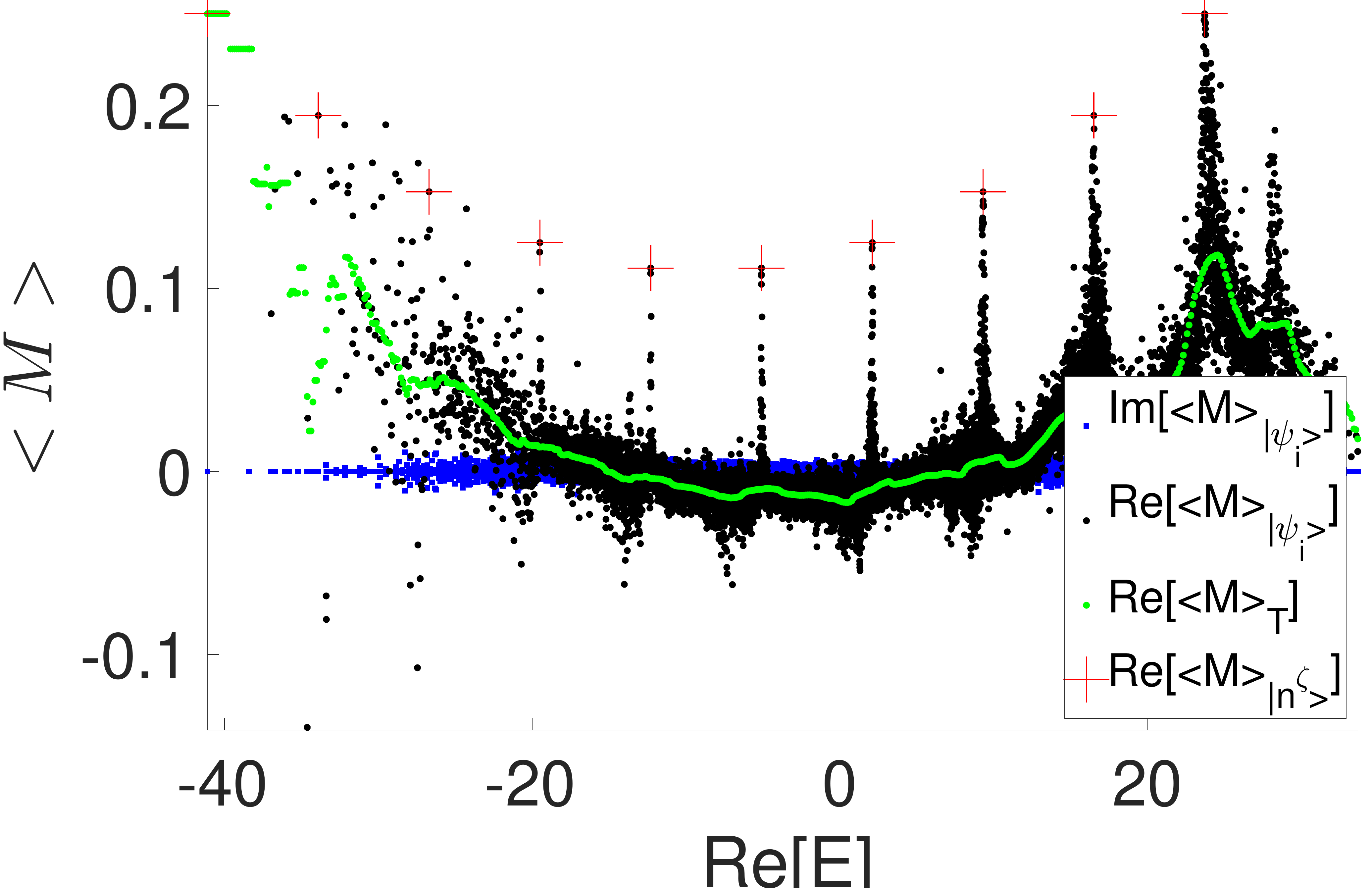}
	\caption{ETH violation by the scar states for the non-Hermitian Hamiltonian \eqref{eq:HnumericsNonHerm}. Shown is the expectation value $\braket{\psi_i|M|\psi_i}$ evaluated for every eigenstate $\psi_i$. The green line is the micro-canonical (window) average.  Left Panel: $M=G_O$ \eqref{eq:Go}, particle number sector $Q=8$. Right Panel: $M = G_U$ \eqref{eq:Gu},  particle number sector $Q=9$, half-filling.}
		\label{fig:ODLRONonHerm}
\end{figure*}

\subsubsection{Quantum chaos}

The level statistics parameters of the hermitian model  $r=0.5306$ ($r^{\rm GOE} = 0.5359$) and the non-hermitian model $r_{nh} =  0.7378$ ($r^{\rm Ginibre} \approx 0.74$) are close to the values of the corresponding random ensembles (defined in Appendix \ref{sec:AppChaos}) and thus indicate that the bulk spectra of both systems are fully chaotic. This is further elaborated by the gap distribution $P(s)$ and by the presence of the "dip-ramp-plateau" structure in the spectral form factor (Fig. \ref{fig:tju_hermSFFandPsG36} and \ref{fig:tju_nonhermPs} ) typical for chaotic systems.

The level spacings distribution, $P(s)$, is the probability density function of the spacings between consecutive eigenvalues. A key feature of the $P(s)$ of random Gaussian matrices is that it decays to zero as $s \rightarrow 0$. This phenomenon is called level repulsion and implies that it is very unlikely for eigenvalues to be nearly identical. We observe level repulsion in both Hermitian (Fig. \ref{fig:tju_hermSFFandPsG36}) and non-Hermitian (Fig. \ref{fig:tju_nonhermPs}) systems studied.

The spectral form factor SFF is usually defined as
\begin{gather}
g(t,\beta) = |Tr(e^{-\beta H - i H t} )|^2 /Tr(e^{-\beta H})^2,
\end{gather}
and gives information about the longer-range correlations of eigenvalues. The main elements of the SFF for a random matrix is a dip ramp plateau structure. The ramp is caused by the repulsion of eigenvalues that are far apart; these eigenvalues are anti-correlated, which is why the ramp is below the plateau. The plateau is a result of generic level repulsion, as degeneracies are unlikely. The ramp and plateau occur at later times and thus probe shorter distances, and these elements are a result of a phenomena known as spectral rigidity. The dip occurs at early times and so it probes larger distances; it is the Fourier transform of the entire spectrum. The probability density function (pdf) of the eigenvalues for a random Hermitian matrix follows a semicircular distribution. This shared property of the pdf leads to a similar slope of the dip in the SFF of the Wigner random ensembles.

In the Hermitian case the dip ramp plateau structure is seen at high enough temperature (Fig. \ref{fig:tju_hermSFFandPsG36}). In the non-Hermitian system the dip ramp plateau structure is best seen if the SFF is calculated for the real parts of the eigenvalues only. In both systems the large magnetic field $\gamma=3.6$ (used to make one of the scar states the ground state) causes the correlations at one corresponding frequency that results in the peak seen in the SFF plot soon after the dip. This peak is absent or much less pronounced for moderate magnetic field of $\gamma=1$ (see Appendix \ref{sec:AppChaos}).

\begin{figure}
	\centering
	\includegraphics[width=\columnwidth]{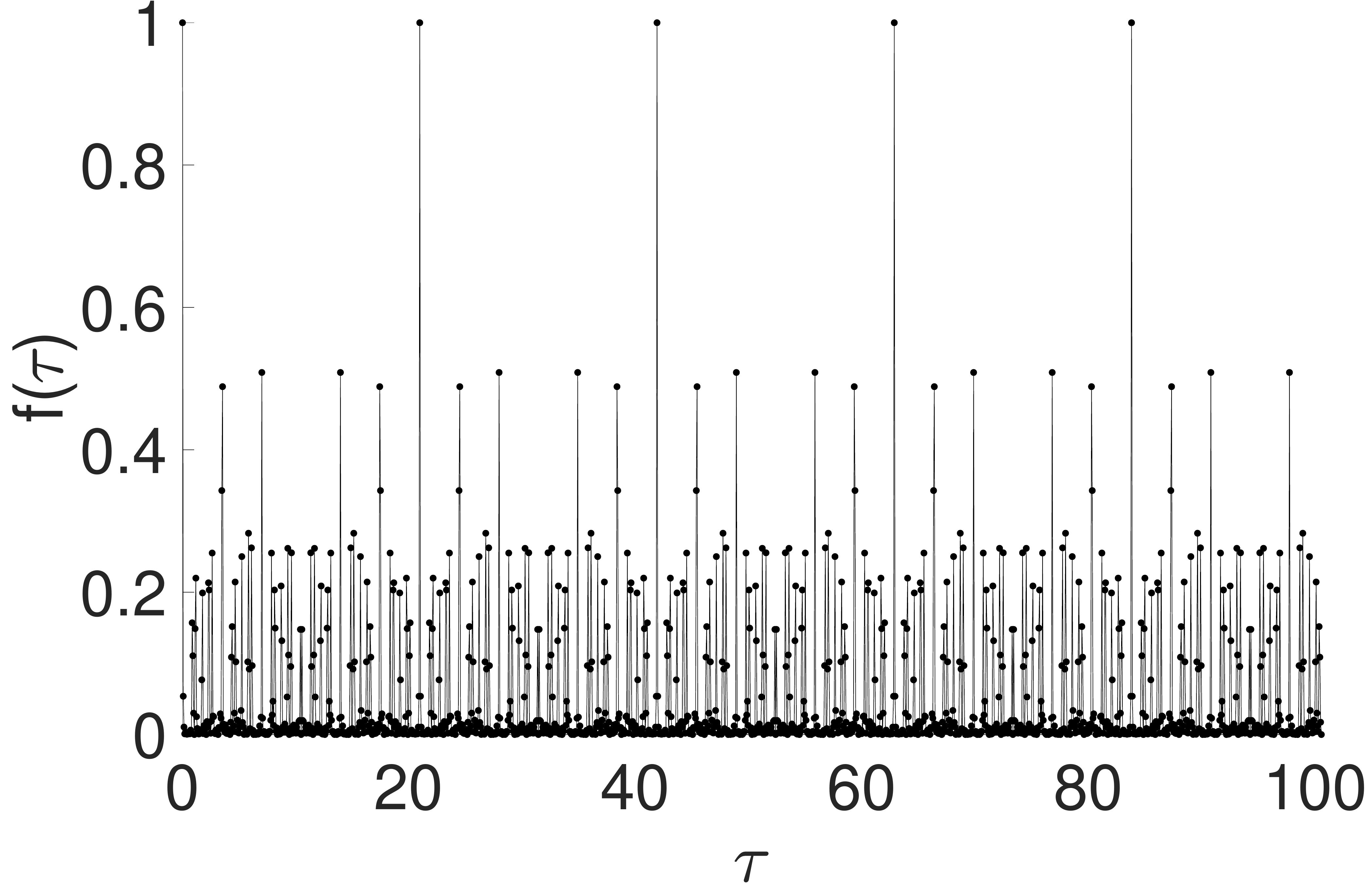}
	\caption{Time dependence of the squared overlap between the initial state (composed of scars only) and a time-evolved state for the non-Hermitian Hamiltonian \eqref{eq:HnumericsNonHerm}. The overlap repeatedly reaches the value of 1 with the period of approximately 20.94 as expected given the energies of the invariant states.}
		\label{fig:nHTimeEvol}
\end{figure}

\subsubsection{Spectrum}

In the spectrum of the non-Hermitian Hamiltonian \eqref{eq:HnumericsNonHerm} shown in Fig. \ref{fig:NHSpec} we observe that all the scar states remain at the real axis and are not effected by the non-Hermitian terms while the majority of the eigenvalues of the non-Hermitian Hamiltonian become complex. This demonstrates the stability of the many-body scar states in suitably designed open systems.

The non-Hermitian spectrum  has a "conjugation symmetry": for every state with energy $a+ib$ there is another state with energy $a-ib$. All the observables measured in any two such states (such as entanglement entropy) are also equal. For this reason we choose to plot such observables as a function of the real part of the energy eigenvalue: $\operatorname{Re} E$.

\subsubsection{Entanglement entropy}

The entanglement entropy at half-filling is shown in Fig. \ref{fig:entropies}. We observe that in both Hermitian and non-Hermitian systems the $\ket{n^\zeta}$ scar states expectedly have entanglement entropy much lower than the rest of the spectrum at corresponding energy (temperature).  This is also true for the $\ket{n^\eta}'$ scars that are located in the respective particle number sectors (not shown). High magnetic field used to make one of the $\ket{n^\zeta}$ states the ground state leads to the emergent structure most clearly seen in Fig. \ref{fig:entropies} for the non-Hermitian case: each total spin sector acquires different energy and starts to form a separate arc as it would be if the total spin was exactly conserved by the Hamiltonian. This additional structure disappears for smaller magnetic field of $\gamma =1$ as seen in the Appendix Fig. \ref{fig:tju_nHermEntrSmallField}.

\subsubsection{ETH violation}

To further demonstrate the violation of strong ETH we evaluate the ``superconducting" 
\begin{gather}
G_O=\braket{c^\dagger_{x_1 y_1 \uparrow} c^\dagger_{x_1 y_1 \downarrow} c_{x_2 y_2 \downarrow} c_{x_2 y_2 \uparrow}}
\label{eq:Go}
\end{gather}
 and ``magnetic" 
 \begin{gather}
 G_U=\braket{c^\dagger_{x_1 y_1 \uparrow} c_{x_1 y_1 \downarrow} c^\dagger_{x_2 y_2 \downarrow} c_{x_2 y_2  \uparrow}}
 \label{eq:Gu}
 \end{gather}
 off-diagonal long-range order (ODLRO) correlators characteristic of the $\ket{n^\eta}'$ \cite{etaPairingYang89} and $\ket{n^\zeta}$ \cite{pakrouski2020GroupInvariantScars} states respectively. We observe (see Fig. \ref{fig:ODLROHerm} and 
\ref{fig:ODLRONonHerm}) that the corresponding expectation values are significantly different in the invariant states relative to the micro-canonical average in both systems which allows us to conclude the invariant states are scars in this system. 

Because of the high symmetry of the invariant states, the above correlators, when evaluated for scars, do not depend on the coordinates $x_1,y_1,x_2,y_2$ \cite{pakrouski2020GroupInvariantScars}. For the numerical evaluation we set the points 1 and 2 to be the most distant points in our system with open boundaries: $(x_1,y_1)=(1,1)$ and $(x_2,y_2)=(3,3)$.
 
Note that while all the $\ket{n^\zeta}$ states are at half-filling ($Q=9$) the particle number of the paired states $\ket{n^\eta}'$ is $2n$ and therefore only one such state is visible for any fixed filling (the data shown is for $Q=8$). 

Very strong magnetic field is present in both systems and couples to the states with non-zero magnetization. This results in the spikes seen in the data for the non-Hermitian system which is apparently more susceptible to the magnetic field.

\begin{figure*}
	\centering
	
	\includegraphics[width=0.49\textwidth]{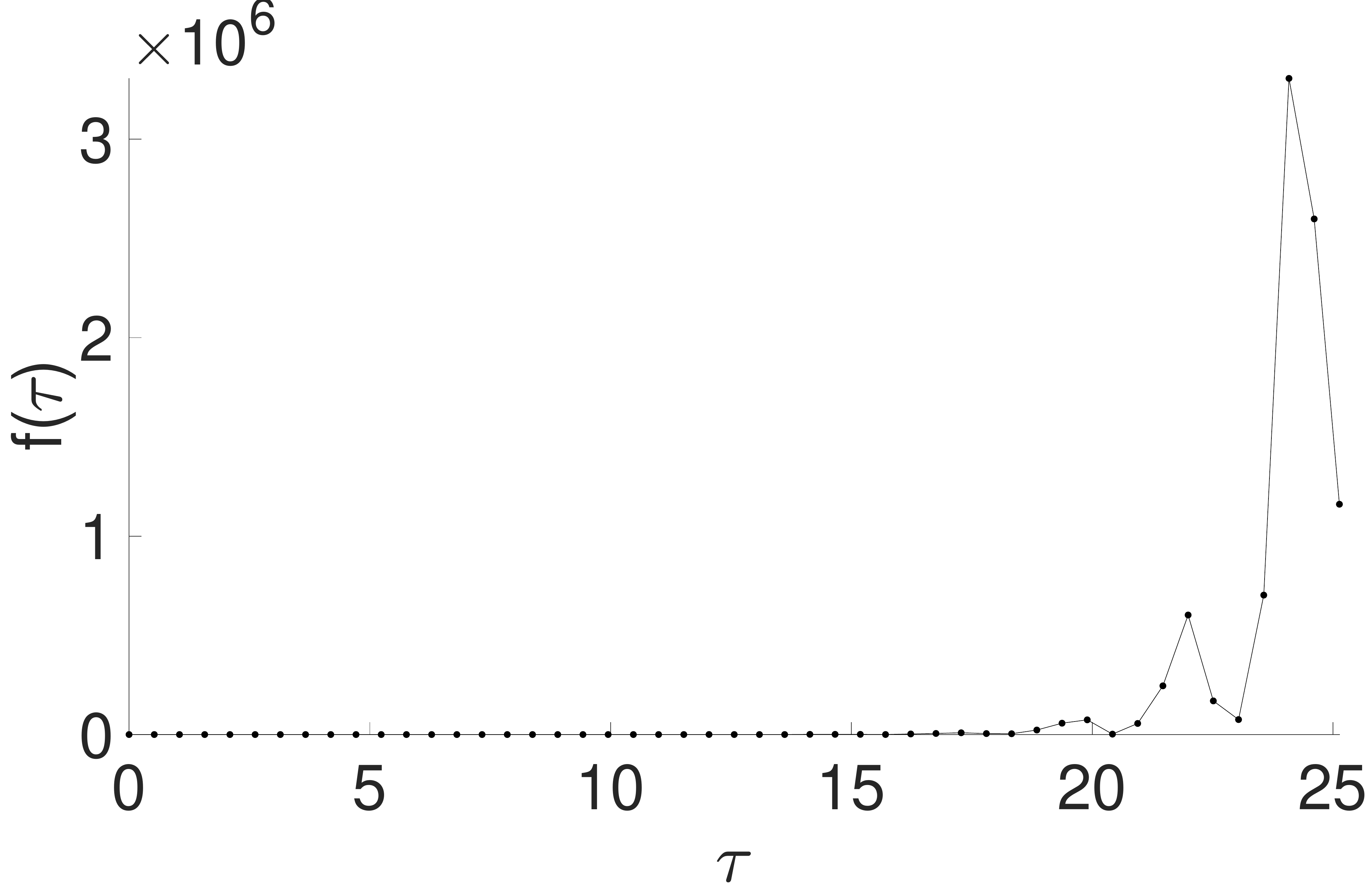} 
	\includegraphics[width=0.49\textwidth]{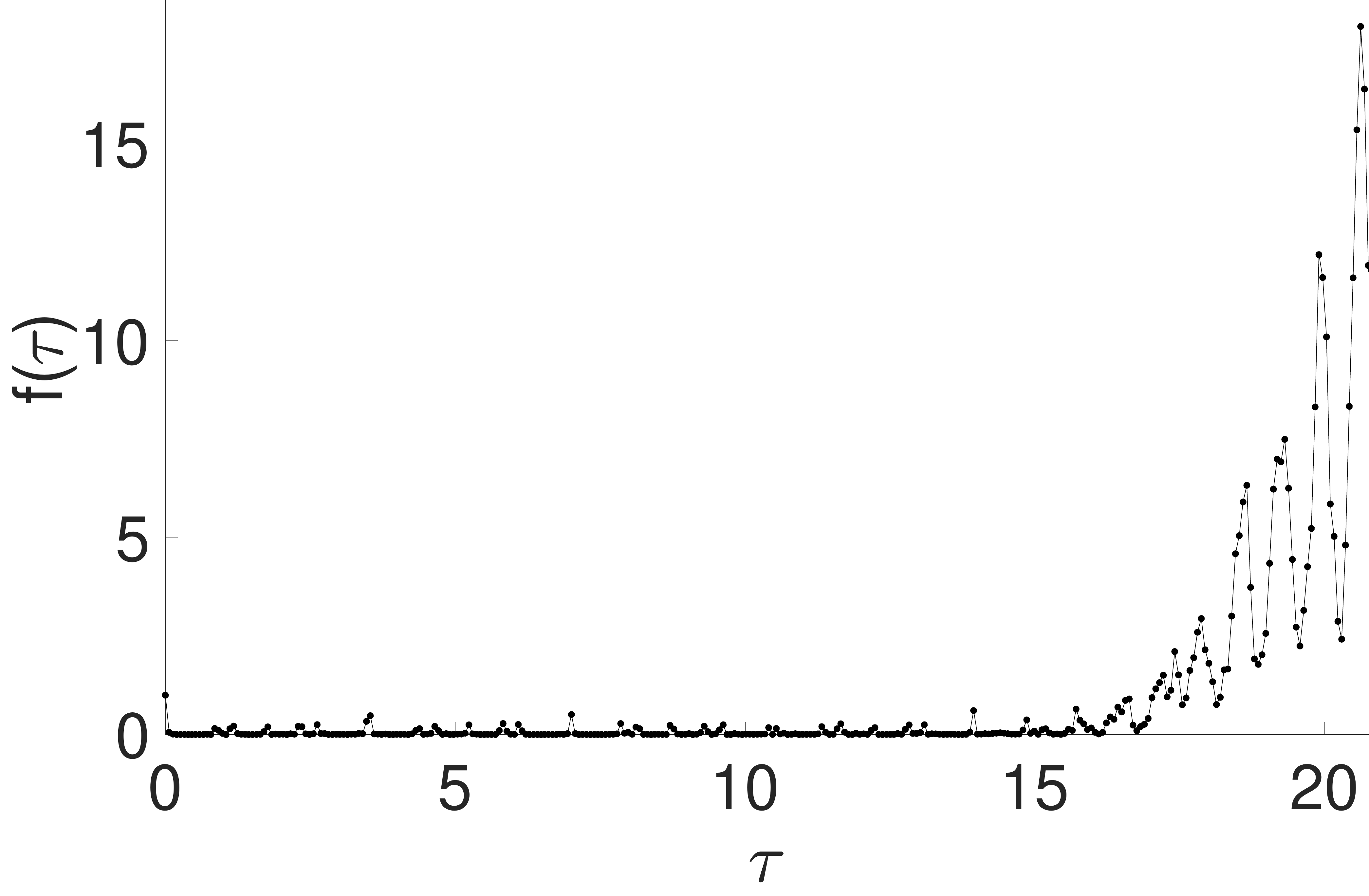}
	
		\caption{Time evolution for the non-Hermitian Hamiltonian \eqref{eq:HnumericsNonHerm}. Squared overlap between the time-evolved and initial states. Left: Initial state is composed of 20 generic states (compare to the time evolution for the initial state composed of 20 scars in Fig. \ref{fig:nHTimeEvol}). Right: In the initial state 1 percent (by weight) of 29 generic states are admixed to 20 scar states (99 percent by weight).}
	\label{fig:initStateTimeEvolNonHerm} 
\end{figure*}

\begin{figure*}
	\centering
	\includegraphics[width=0.49\textwidth]{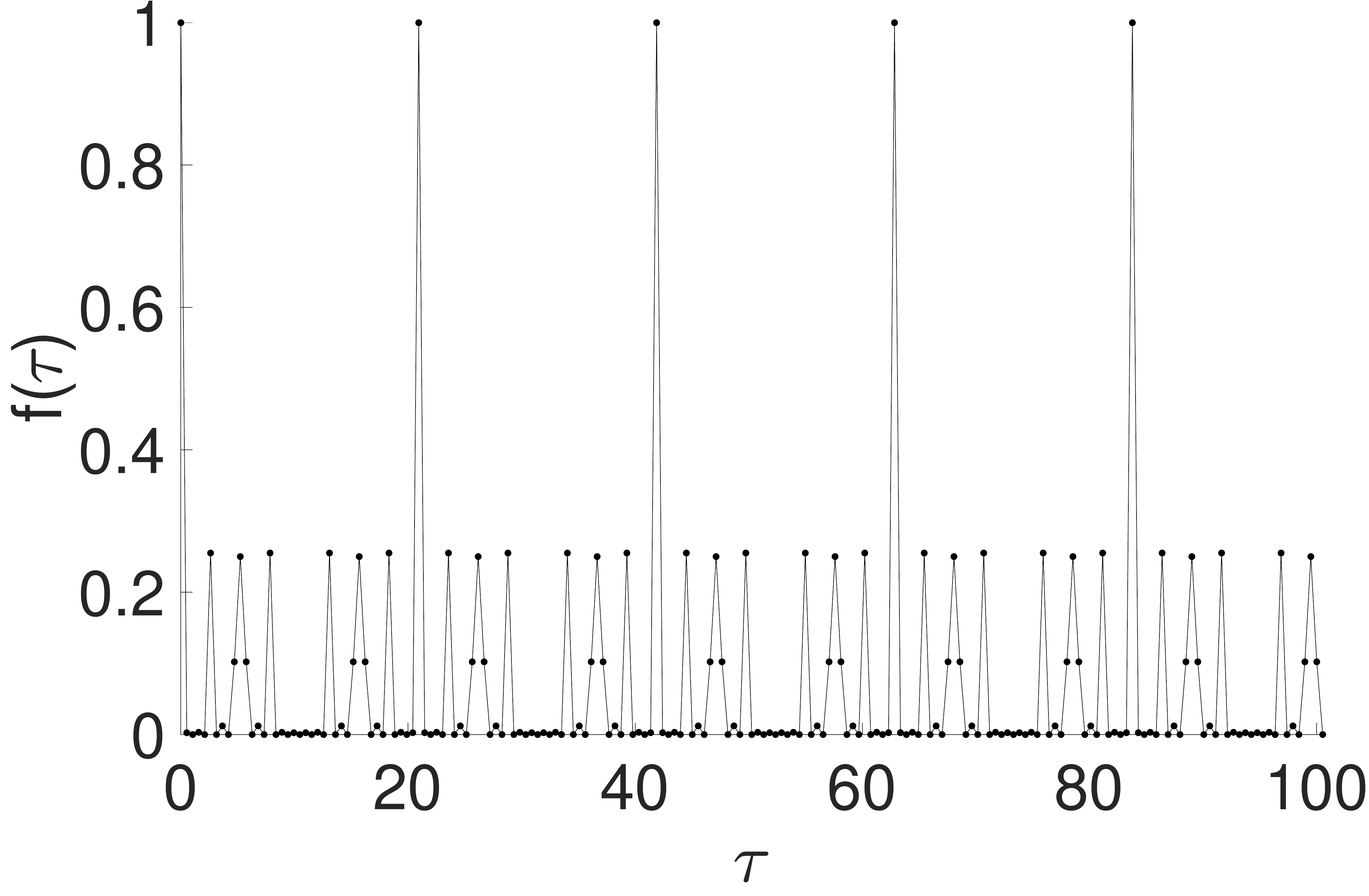}
	\includegraphics[width=0.49\textwidth]{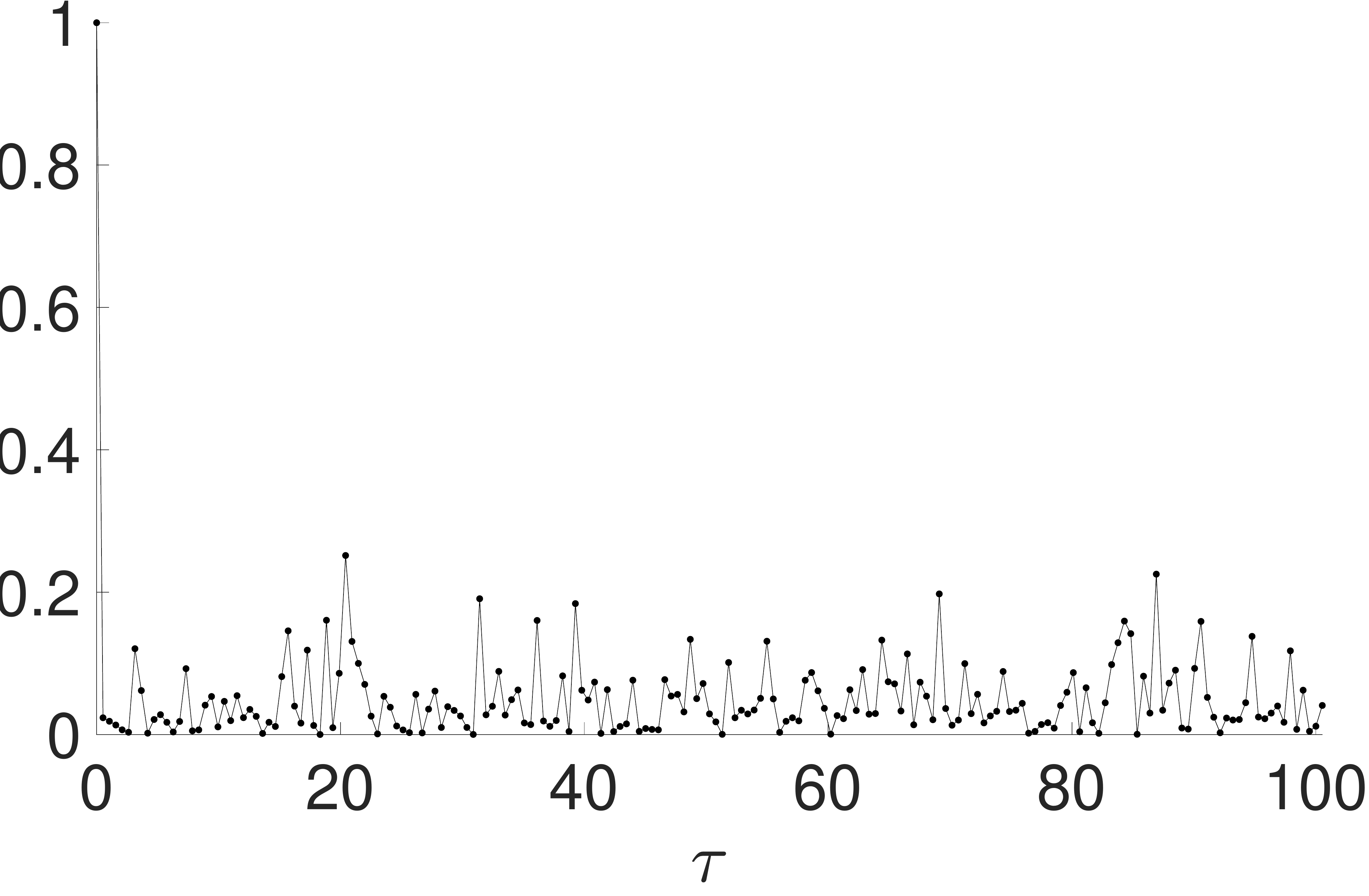}
	
\caption{Time evolution for the Hermitian Hamiltonian \eqref{eq:HnumericsHerm}. Squared overlap between the time-evolved and initial states. Left: Initial state is composed of 20 scars. Right: Initial state is composed of 20 generic states.	}	\label{fig:initStateTimeEvolHerm} 
\end{figure*}

\subsubsection{Time evolution and revivals}

One of the most striking and counter-intuitive features of the scar states in the non-Hermitian Hamiltonian \eqref{eq:HnumericsNonHerm} is the stable and coherent time evolution of the scar subspace shown in Fig. \ref{fig:nHTimeEvol}. The system is initialised to a state that is a uniform mix of all the $2N+2=20$ scar states present in the system. As shown in Fig. \ref{fig:initStateTimeEvolNonHerm} this state is coming back to itself exactly after the time intervals $2\pi/\omega \approx20.94$, where $\omega=0.3$ is the greatest common divisor of all the gaps between the scar states according to \eqref{eq:nonuEnInNum}. The norm of the state is preserved throughout, although the system is open (Hamiltonian is non-Hermitian).

For the initial state composed of generic eigenstates the imaginary components of the eigenvalues lead to the probability density quickly flowing into the effectively open system. As can be seen in the left panel of Fig. \ref{fig:initStateTimeEvolNonHerm}, the norm of the time-evolved state and correspondingly the overlap explode already over one revivals period.

A similar phenomenon is observed also if only 1 percent (by weight) of generic, complex eigenstates is admixed to the initial state dominated (99 percent by weight) by scars. The initially small imaginary components obtain exponential amplification with time, however a few periods of the revivals with the same period as in other systems can still be observed as shown in the right panel of Fig. \ref{fig:initStateTimeEvolNonHerm}.

As expected, starting from a mix of scar states, stable revivals are also observed for the Hermitian Hamiltonian (see Fig. \ref{fig:initStateTimeEvolHerm}). In contrast, the information initially stored in the generic states quickly dissipates through thermalisation (right panel of Fig. \ref{fig:initStateTimeEvolHerm}).

\section{Discussion}

We have shown explicitly that the Hamiltonians of some well-known models (Table \ref{tab:knownModelsAsOT}) are of the $H_0+OT$ form; therefore, they support the group-invariant scars for any coupling constants and without a need for fine-tuning. A large number of other models can be built by combining the terms we list as the group generators in Sec. \ref{sec:nOnU}. Of particular interest are the ``superconducting" terms in \eqref{hop:T+} and the spin-orbit coupled hopping terms in \eqref{newgen} and \eqref{hop:Ttilalpha}. It will be very interesting to explore the interplay of the weak ergodicity breaking and the superconductivity or topology by studying the models built of such terms.

Another possible generalization of our approach is to consider discrete groups $G$ instead of Lie algebras. In this case we should use $T=g-1$ with $H=H_O + O T$, where $g \in G$ and $g^{-1} H_0 g = H_0$.

From the quantum computation perspective, the group-invariant scar states we consider are analogous to a "decoherence-free" subspace that can be used to reduce noise in the universal quantum computation \cite{ZanardiNoiselessQuantumCodes,DecohFreeQComputation1998Chuang}. Thus, another interesting direction for future work is the development of specific protocols that would enable more robust quantum computation.

Many of the models we are considering are among the most commonly and widely used in condensed matter physics. Therefore there is a great potential for studying these models and the weak ergodicity breaking effects therein in cold atoms quantum simulations, and also for identifying materials that could host this phenomenon. We do describe conditions under which the ground state is a many-body scar or the full low-energy subspace consists solely of scars, which should further facilitate the experimental explorations.

While in this work we considered the case of spin-1/2 fermionic systems, the general approach of Ref. \cite{pakrouski2020GroupInvariantScars} can be readily applied to the systems with larger number of fermionic species per site and richer structure of the singlet subspace. Such examples will be considered in our subsequent work.

Finally, we have demonstrated that the invariant scar subspace continues to undergo stable, coherent time evolution in a class of suitably designed open systems with non-Hermitian Hamiltonian. This greatly expands the realm of weak ergodicity breaking phenomena and will hopefully inspire new theoretical and experimental studies. In particular, it would be interesting to study the relation of the degenerate group-invariant scar subspaces in the non-Hermitian systems (e.g. in \eqref{eq:nonuEnInNum}, $B_y=0$ for $\ket{n^\zeta}$ or $\mu=U/2$ for $\ket{n^\eta}'$) to the ``exceptional points," the degeneracies in the non-Hermitian operators that lead to numerous exotic phenomena and are being intensively studied currently \cite{Berry2004ExcPTalk,Heiss2012ExcP,Mirieaar2019ExcPScience,Soriente2021}. 

\section{Acknowledgements}
We thank A. Bernevig, A. Dymarsky, D. Haldane, S. Moudgalya, B. Bu{\v c}a and A. Prem for useful discussions and A. Dymarsky for his valuable comments on a draft of this paper. I.R.K. is very grateful to the Institute for Advanced Study for hosting him during his sabbatical leave. 
The simulations presented in this work were performed on computational resources managed and supported by Princeton's Institute for Computational Science $\&$ Engineering and OIT Research Computing.
This research was supported in part by the US NSF under Grant No.~PHY-1914860. 
K.P. was also supported by the Swiss National Science Foundation through the Early Postdoc.Mobility Grant No. P2EZP2$\_$172168 and by DOE grant No. de-sc0002140.
F.K.P. is currently a Simons Junior Fellow at NYU and supported by a grant 855325FP from the Simons Foundation.

\appendix

\section{Derivation of the $H_0+OT$ form of the Hamiltonian}
\label{sec:SMDeriveOT}
Throughout this section we will use the following definitions: $\nu=\frac{Q}{2N}$ is the filling fraction; $N^{nn}_1$ and $N^{nn}_2$ are the numbers of nearest- and next-to-nearest-neighbor pairs in a particular lattice.

\begin{table}[t!]

	\begin{center}
		\begin{tabular}{|c|c|c|c|c|c|c|}
			\hline
		$\ket{}$ 		&	$\sum_i H^{Hub}_i$ 		& $\sum_i M_i$  		& $Q$  	& $\vec{S}_i \cdot \vec{S}_j $  	& $\sum_i K_i$		& $\sum_i K_i^2$ \\
		\hline
		$\ket{n^\zeta}$ 	& 0 						&  $N-2n$				& $N$ 			& $\frac14$ 					& 0				& 0 \\
		\hline
		$\ket{n^\eta}$ 	& $n$					& 0 					& $2n$ 			& 0  						& $2n-N$			& $N$ \\
		\hline
		$\ket{n^\eta}'$ 	& $n$					& 0 					& $2n$ 			& 0  						& $2n-N$			& $N$ \\
		\hline
	\end{tabular} 
	\end{center}
	\caption{ The action of various operators on the invariant states. Summation $\sum_i$ goes over all the lattice sites. Lattice can be arbitrary, the $\ket{n^\eta}'$ states are defined however on a bipartite lattice only.
	}	
\label{tab:summedGenOnStates}
\end{table}

\subsection{Hubbard model \label{sec:SMHubbardModelAsOT}}

The Hubbard model Hamiltonian reads 
\begin{gather*}
\label{eq:HHubMod}
H^{Hub} = t \sum\limits_{\braket{i,j},\sigma} \left[c^\dagger_{j\sigma} c_{i\sigma} + c^\dagger_{i\sigma} c_{j\sigma}\right] - \mu Q 
+ U \sum\limits_i  H^{Hub}_i \ ,
\end{gather*}
where $i$ and $j$ label sites of an arbitrary lattice in arbitrary dimension, $\braket{i,j}$ stands for the nearest neighbors, and $t$ is a negative real number.

Substituting our result \eqref{eq:hubi} we obtain the $H_0+OT$ form in terms of the generators of U$(N)$,
\begin{gather}
\label{eq:HubDecompZeta}
H^{Hub} = Q\left(\frac{U}{2} -\mu\right) - U \frac{N}{2} + t \sum_{\braket{i,j}}  T'_{\braket{i,j}}  + \frac{U}{2} \sum\limits_i K^2_i\ ,
\end{gather}
where the hopping term is a generator of SU$(N)$ as a special case of $T_{ij}$ \eqref{hop:T}.
On a bipartite lattice this hopping term coincides with \eqref{hop:T'} is therefore a generator of $\widetilde{\rm SU}(N)'$. Using \eqref{eq:KiThroughMi}  we obtain the $H_0+OT$ decomposition for the group $\widetilde{\rm SU}(N)'$:
\begin{gather}
H^{Hub} =Q \left(\frac{U}{2} -\mu\right) +  t \sum_{\braket{i,j}}  T'_{\braket{i,j}}  - \frac{U}{2} \sum\limits_i M_i^2.
\label{eq:HubDecompEta}
\end{gather}
Comparing the two expressions \eqref{eq:HubDecompZeta} and \eqref{eq:HubDecompEta}, we notice that they are similar except that the constant $U\cdot N/2$ belongs to $H_0$ in the first and to $OT$ in the second equation (as a consequence of \eqref{eq:KiThroughMi}).

Only the $\ket{n^\zeta}$ states are scars for the Hubbard model on an arbitrary lattice.
On a bipartite lattice, because the $H_0+OT$ decompositions are possible w.r.t. two different groups SU$(N)$ and $\widetilde{\rm SU}(N)'$, the singlets of these groups $\ket{n^\zeta}$ and $\ket{n^\eta}'$ together form the $2N+2$-dimensional scar subspace.
In the pure Hubbard model the $\ket{n^\zeta}$ subspace ($Q=N$) is degenerate with energy (see \eqref{eq:HubDecompZeta})
\begin{gather}
H^{Hub}\ket{n^\zeta} =- \mu N \ket{n^\zeta}
\end{gather}
while the $\eta$-paired states $\ket{n^\eta}'$ ($Q=2n$) are split equidistantly according to their particle number (see \eqref{eq:HubDecompEta})
\begin{gather}
H^{Hub}\ket{n^\eta}' = (U-2\mu) n \ket{n^\eta}'.
\end{gather}
The states $\ket{n^\eta}$ are not eigenstates of the Hubbard model; they are mixed by the hopping terms.

\subsection{Heisenberg model  \label{sec:SMHeisenberg} }

The spin-1/2 Heisenberg Hamiltonian is
\begin{gather}
\label{eq:HeisMod}
H_{\rm Heisenberg} = J \sum_{\braket{ij}} \vec{S}_i \cdot \vec{S}_j .
\end{gather}
The constraint $\vec S_i \cdot \vec S_i=3/4$
translates into the requirement of half-filling on each site: $n_i=1$. 
This sector of Hilbert space contains the $\ket{n^\zeta}$ states and a single, half-filled $\ket{n^\eta}/\ket{n^\eta}'$ state.
When we restrict to the locally half-filled subspace $K_i=0$, we get from \eqref{eq:HHeisij}
\begin{gather}
\label{eq:heisHalfFill}
	\vec{S}_i \cdot \vec{S}_j= \frac{1}{4} -  \frac12 E_{ij} E_{ji}
\end{gather}

and the $H_0+OT$ form (w.r.t. SU$(N)$) is
\begin{gather}
H_{\rm Heisenberg} = \frac{J}{4}N^{nn}_1 + \frac{J}{2} \sum\limits_{\braket{ij}} E_{ij} E_{ji},
\end{gather}
where $N^{nn}_1$ is the number of nearest-neighbour links on a given lattice and $E_{ij}$ are the SU$(N)$ generators.

Therefore the U$(N)$-invariant states $\ket{n^\zeta}$ are degenerate eigenstates of the Heisenberg model with the energy
\begin{gather}
E^{Heis}_{\ket{n^\zeta}} = \frac{J}{4}N^{nn}_1.
\end{gather}

\subsection{Haldane-Shastry model}

The isotropic ($\Delta=0$) Hamiltonian \cite{HaldaneOfHaldaneShastry} of the 1D Haldane-Shastry model \cite{HaldaneOfHaldaneShastry,ShastryOfHaldaneShastry} reads
\begin{gather}
H^{HS} = \sum_{n<n'} J(n-n') \vec{S}_n \cdot \vec{S}_{n'},
\end{gather}
with $J(n-n') = \frac{\pi^2}{N^2 \sin^2\left(\frac{\pi(n-n')}{N}\right)}$.

We have seen that each of such terms acts on $\ket{n^\zeta}$ states with a constant (Sec. \ref{sec:HeisInt} and Eq. \eqref{eq:HHeisij}): $\vec{S}_i \cdot \vec{S}_j \ket{n^\zeta} = \frac{1}{4}\ket{n^\zeta}$.

Therefore all the $\ket{n^\zeta}$ states are degenerate in the isotropic Haldane-Shastry model, are scars and have the energy 
\begin{gather}
E^{HS}_{n_U} = \frac{\pi^2}{4N^2} \sum_{k} \frac{1}{ \sin^2 \left(\frac{\pi k}{N}\right)} = \frac{\pi^2}{4 N^2} \frac{N^3-N}{6},
\end{gather}
which is always integer in units of $\frac{\pi^2}{4 N^2}$. We note that Ref. \cite{HaldaneOfHaldaneShastry} does mention the existence of such integer-energy states.\footnote{The precise statement in Ref. \cite{HaldaneOfHaldaneShastry} is
``For the isotropic model, the numerical study reveals a
surprising fact: States are grouped into highly degenerate
supermultiplets, and at every value of the crystal
momentum and parity, every energy level is contained in
the set derived from states with real pseudomomenta,
and the energies in units of $\frac{\pi^2}{4 N^2}$ are all integers."}

Both families of paired states are exactly annihilated by the Haldane-Shastry Hamiltonian as a special case of the Heisenberg interaction.

\subsection{$J_1-J_2$ model}
$J_1$-$J_2$ model is the Heisenberg interaction between nearest-neighbours ($J_1$) and next-nearest-neighbours ($J_2$):
\begin{gather}
H^{J_1J_2} = J_1 \sum_{\braket{ij}} \vec{S}_i \cdot \vec{S}_j  + J_2 \sum_{\braket{\braket{kl}}} \vec{S}_k \cdot \vec{S}_l.
\end{gather}

Using Eq. \eqref{eq:HHeisij} we write
\begin{gather}
H^{J_1J_2} = J_1 \sum_{\braket{ij}} H^{Heis}_{ij}  + J_2 \sum_{\braket{\braket{kl}}} H^{Heis}_{kl} = \notag \\
J_1 \sum_{\braket{ij}} \left( C_{ij} + \frac14 \right) + J_2 \sum_{\braket{\braket{kl}}} \left( C_{kl} + \frac14 \right) = \notag \\
\frac14\left(J_1 N^{nn}_1 + J_2 N^{nn}_2\right)  
+ J_1 \sum_{\braket{ij}} C_{ij} + J_2 \sum_{\braket{\braket{kl}}} C_{kl},
\end{gather}
where the SU$(N)$-generators $C_{ij}$ are defined in Eq. \eqref{eq:THeis}.

It follows that $H^{J_1J_2}\ket{n^\zeta}=\frac{1}{4} \left(J_1 N^{nn}_1 + J_2 N^{nn}_2 \right) \ket{n^\zeta}$, while $H^{J_1J_2}$ annihilates the paired states $\ket{n^\eta}$ and $\ket{n^\eta}'$ just as each of its terms does individually. Therefore all three families  decouple in the $J_1$-$J_2$ model and are scars.

We also note that the above results directly apply to the Majumdar-Ghosh model which is a special case with $J_1=2J_2$.

\subsection{Hirsch model}

\subsubsection{The ``reduced" version \cite{Hirsch1989Reduced,2020MarkMotrEtaPairHub}}

The ``reduced" Hirsch model for which the $\ket{n^\eta}'$ states were shown to be scars in Ref. \cite{2020MarkMotrEtaPairHub} reads

\begin{gather}
    H^{\rm Hir}_{\rm r} =  -\sum_{\langle ij \rangle, \sigma} \left[t - X (n_{i,-\sigma} + n_{j,-\sigma}) \right]  \left(c^\dagger_{i,\sigma} c_{j,\sigma} + h.c. \right) \notag \\
+ U \sum_j n_{j, \uparrow} n_{j,\downarrow} - \mu Q,
\label{eq:Hirschmodel}
\end{gather}
where $X$ is a real number and $-\sigma$ is the opposite direction of spin $\sigma$ or equivalently \cite{2020MarkMotrEtaPairHub}

\begin{gather}
    H^{\rm Hir}_{\rm r} =  -\sum_{\langle ij \rangle} \left[t - X (n_{i} + n_{j}-1) \right]  \sum_\sigma  \left(c^\dagger_{i,\sigma} c_{j,\sigma} + h.c. \right) \notag \\
+ U \sum_j n_{j, \uparrow} n_{j,\downarrow} - \mu  Q.
\label{eq:HirschmodelMinus1}
\end{gather}

Defining
\begin{gather}
O^{HR}_{ij} = -\left[t - X (n_{i} + n_{j}-1) \right] ,
\end{gather}
noticing that the terms $T'_{\braket{ij}}$ are generators of both $\widetilde{\rm SU}(N)'$ and SU$(N)$ (see \eqref{hop:T'}) and using Eq. \eqref{eq:hubi} for $n_{j, \uparrow} n_{j,\downarrow}$ we obtain the $H_0+OT$ form for the model \eqref{eq:HirschmodelMinus1} with respect to the group SU$(N)$ ($\ket{n^\zeta}$ scars)

 \begin{gather}
 H^{\rm Hir}_{\rm r} =  \sum_{\langle ij \rangle} O^{HR}_{ij} T'_{\braket{ij}}+ \frac{U}{2}\sum_j K_j^2 + Q \left(\frac{U}{2} -\mu\right)  -\frac{UN}{2}
 \end{gather}
and with respect to the group $\widetilde{\rm SU}(N)'$ ($\ket{n^\eta}'$ scars)

 \begin{gather}
 H^{\rm Hir}_{\rm r} =  \sum_{\langle ij \rangle} O^{HR}_{ij} T'_{\braket{ij}} + Q \left(\frac{U}{2} -\mu\right) - \frac{U}{2} \sum\limits_i M_i^2
 \end{gather}

On a generic lattice $T'_{\braket{ij}}$ is a generator of SU$(N)$ and the half-filled SU$(N)$-invariant $\ket{n^\zeta}$ states become scars with the energies,
\begin{gather}
H^{\rm Hir}_{\rm r} \ket{n^\zeta} =- \mu N \ket{n^\zeta}.
\end{gather}
On a bipartite lattice the nearest-neighbour hopping  $T'_{\braket{ij}}$ is a generator of $\widetilde{\rm SU}(N)'$. Therefore in addition to the $\ket{n^\zeta}$ the $\ket{n^\eta}'$ states become scars with the energies
\begin{gather}
H^{\rm Hir}_{\rm r}\ket{n^\eta}' = \left(U -2 \mu \right) n \ket{n^\eta}'.
\end{gather}

\subsubsection{The full version from Ref. \cite{HIRSCHModelOriginal}}

The Hirsch model in its original formulation (Eq. 6 in Ref. \cite{HIRSCHModelOriginal}, see also \cite{Micnas1989SortOfHirsch}) contains an additional density-density term but is missing the chemical potential
\begin{gather}
H^{\rm Hir}= -t_0 \sum_{\langle ij \rangle, \sigma} \left(c^\dagger_{i,\sigma} c_{j,\sigma} + {\rm h.c.} \right) + U \sum_i H^{Hub}_i + \notag \\
+ V \sum_{\langle ij \rangle}n_i n_j +\Delta t \sum_{\langle ij \rangle, \sigma} \left(c^\dagger_{i,\sigma} c_{j,\sigma} +{\rm h.c.} \right)  (n_{i,-\sigma} + n_{j,-\sigma}).
\label{eq:Hirschmodel}
\end{gather}

Defining
\begin{gather}
O^{HF}_{ij} = \left[-t_0 + \Delta t (n_{i} + n_{j}-1) \right] 
\end{gather}
and using \eqref{eq:densDens} for density-density we bring the full Hirsch model to the $H_0+OT$ form w.r.t. the group SU$(N)$
\begin{gather}
 H^{\rm Hir} =\frac{U}{2}(Q-N)  + V N^{nn}_1 +  \\
+\sum_{\langle ij \rangle} O^{HF}_{ij} T'_{\braket{ij}}   + \frac{U}{2}\sum_j K^2_j + V \sum_{\langle ij \rangle} \left(K_i K_j+K_i+K_j\right). \notag
\end{gather}

We have seen in Sec. \ref{sec:densityDensityInter} that the generic density-density interaction can be written in terms of the SU$(N)$ generators but doesn't have the paired states as eigenstates.

Therefore only $\ket{n^\zeta}$ states with energies 
 \begin{gather}
H^{\rm Hir} \ket{n^\zeta}  = V N^{nn}_1 \ket{n^\zeta}
 \end{gather}
 are scars in the full Hirsch model \cite{HIRSCHModelOriginal}.

\subsection{Haldane-Hubbard model \label{sec:HHmodel}}

The spin-full Haldane-Hubbard model is \cite{HaldaneModel1988,HHFeng2011,HH2018PhysicaB}
\begin{gather}
\label{eq:HHubMod}
H^{HH} = t_1 \sum\limits_{\braket{i,j},\sigma} \left[c^\dagger_{j\sigma} c_{i\sigma} + c^\dagger_{i\sigma} c_{j\sigma}\right] - \mu Q
+ \notag\\ 
+ U \sum\limits_i  H^{Hub}_i  + t_2 \sum_{\braket{\braket{kl}} ,\sigma} \left[ e^{i\Phi_{kl}} c^\dagger_{k\sigma} c_{l\sigma} + e^{-i\Phi_{kl}} c^\dagger_{l\sigma} c_{k\sigma}\right] ,
\end{gather}
and can be defined on any bipartite lattice and in particular on the 2D honeycomb lattice of graphene.

The model is obtained by adding the complex-amplitude next-nearest-neigbour hopping to the Hubbard model which means the $H_0+OT$ decomposition only differs from Hubbard model by the addition of the $t_2$ hopping to the $OT$ (see Tab. \ref{tab:knownModelsAsOT}).

In the most general case the $t_2$ hopping terms  are the generators of SU$(N)$ and the model possesses only one scar family $\ket{n^\zeta}$ with the corresponding $H_0+OT$ decomposition
\begin{gather}
\label{eq:HHubModOTNU}
H^{HH} = Q \left(\frac{U}{2} -\mu\right)  - U \frac{N}{2} + 
t_1 \sum_{\braket{i,j}}  T'_{\braket{i,j}}  +  \notag \\\frac{U}{2} \sum\limits_i K^2_i  + t_2 \sum_{\braket{\braket{kl}} ,\sigma} \left[ e^{i\Phi_{kl}} c^\dagger_{k\sigma} c_{l\sigma} + e^{-i\Phi_{kl}} c^\dagger_{l\sigma} c_{k\sigma}\right] .
\end{gather}

For $\Phi_{kl} = \pi/2$ the amplitude of the $t_2$ hopping becomes purely imaginary and because the lattice in the Haldane-Hubbard model is always bipartite the hopping $t_1+t_2$ is a generator of $\widetilde{\rm SU}(N)'$ (see Sec \ref{sec:SMrealHopAmpl}) and the $\ket{n^\eta}'$ family of scars is added while the corresponding $H_0+OT$ decomposition reads 
\begin{gather}
H^{HH}(\Phi_{kl} = \pi/2) = Q \left(\frac{U}{2} -\mu\right) +
 t_1 \sum_{\braket{i,j}}  T'_{\braket{i,j}}  + \notag \\ +i t_2 \sum_{\braket{\braket{kl}} ,\sigma} \left[  c^\dagger_{k\sigma} c_{l\sigma} - c^\dagger_{l\sigma} c_{k\sigma}\right]  - \frac{U}{2} \sum\limits_i M_i^2.
\label{eq:HHubModOTNOp}
\end{gather}

If in addition $t_1=0$ then the remaining $t_2$ hopping alone is a generator (see Eq. \eqref{hop:tilT}) of SO$(N)$ which is a subgroup of both $\widetilde{\rm SU}(N)'$ and $\widetilde{\rm SU}(N)$ and the third scar family of $\widetilde{\rm SU}(N)\times \rm SU(2)_{spin}$-invariant states $\ket{n^\eta}$ is added. The corresponding $H_0+OT$ decomposition is
\begin{gather}
H^{HH}(\Phi_{kl} = \pi/2; t_1=0) = Q \left(\frac{U}{2} -\mu\right) +\notag \\
+ i t_2 \sum_{\braket{\braket{kl}} ,\sigma} \left[  c^\dagger_{k\sigma} c_{l\sigma} - c^\dagger_{l\sigma} c_{k\sigma}\right]  - \frac{U}{2} \sum\limits_i M_i^2.
\label{eq:HHubModOTNO}
\end{gather}

\section{Quantum chaos \label{sec:AppChaos}}

\begin{figure*}
	\centering
	\includegraphics[width=0.49\textwidth]{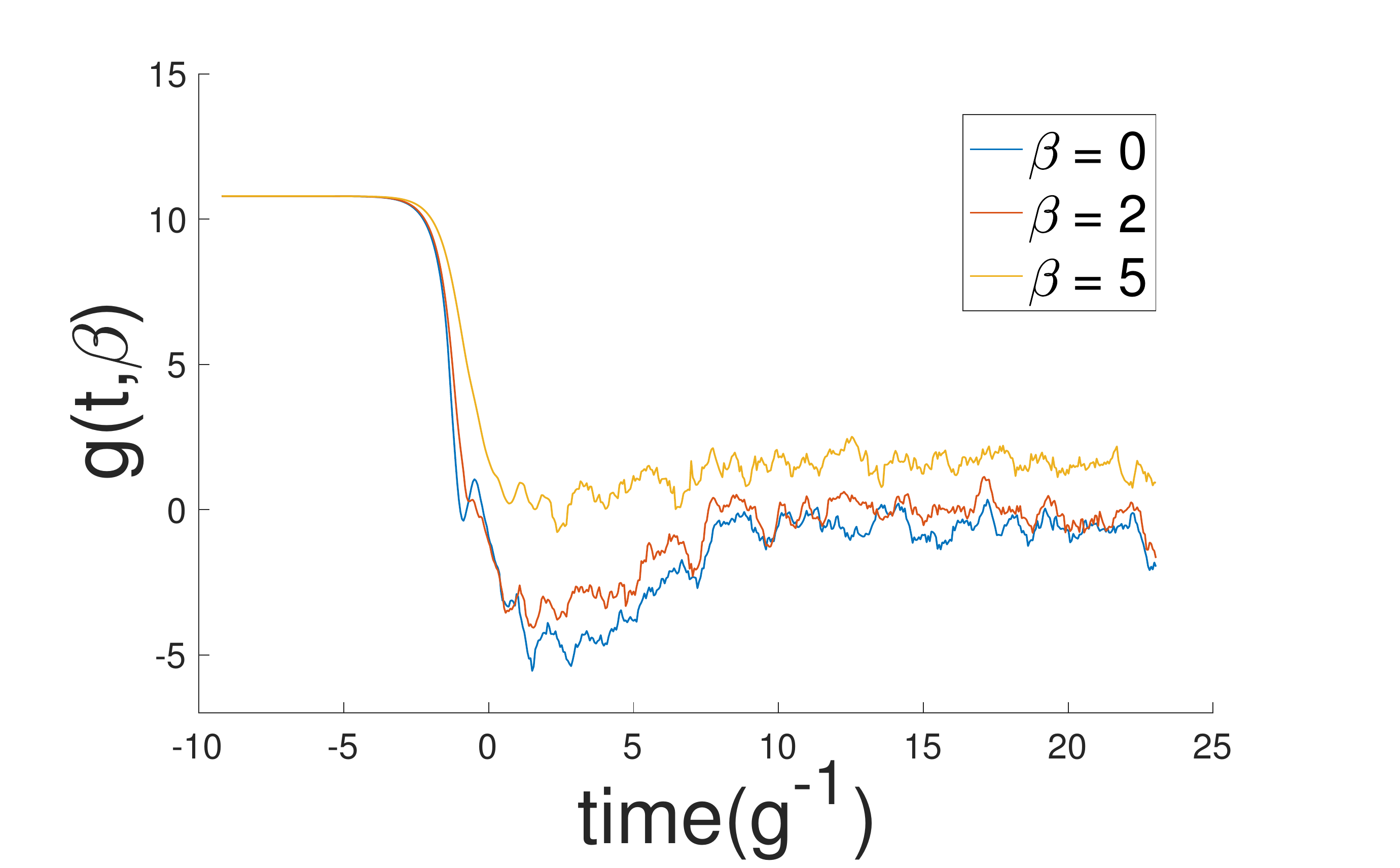}
	\includegraphics[width=0.49\textwidth]{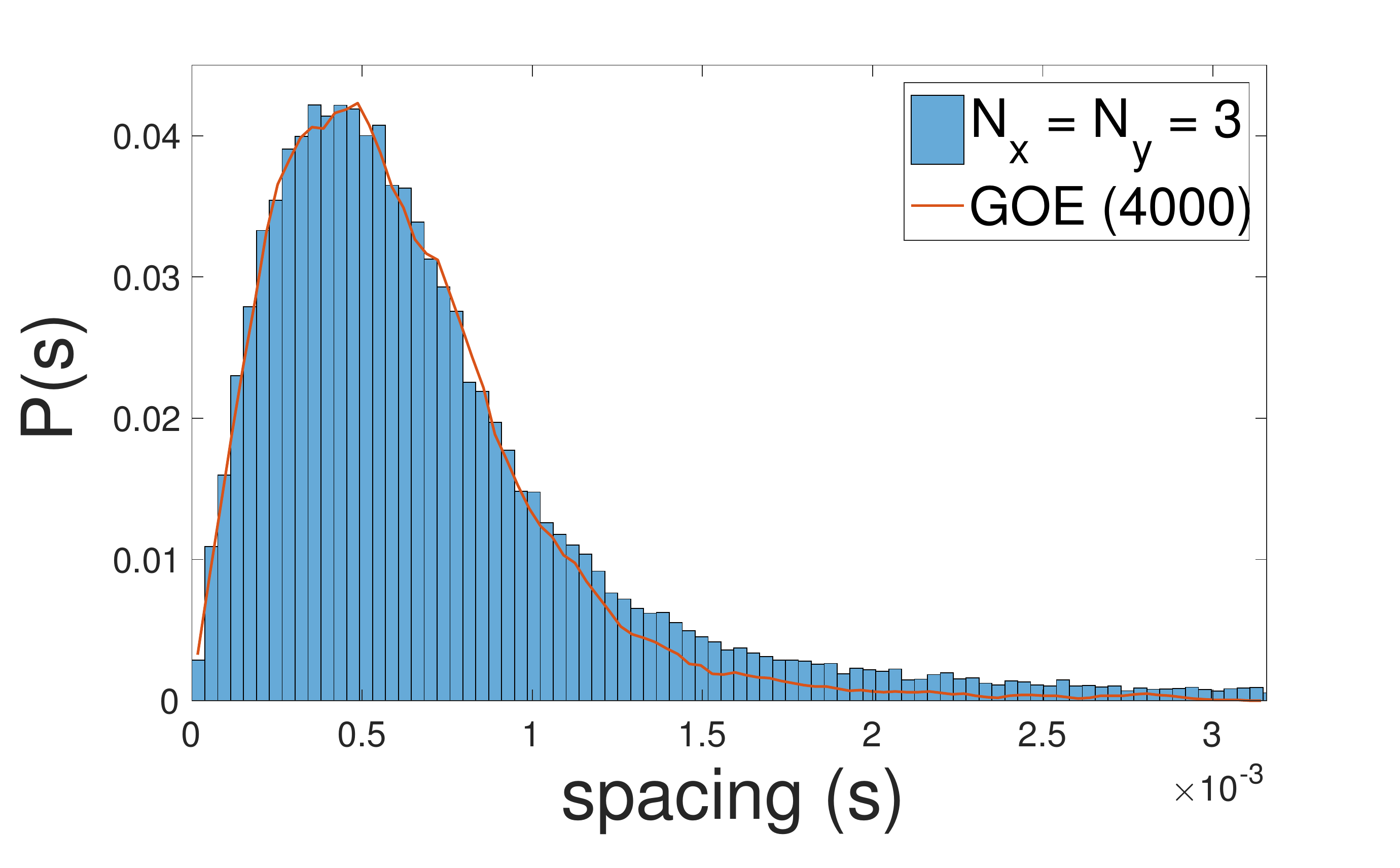}
	\caption{Quantum chaos for Hermitian $tJU$ \eqref{eq:HnumericsHerm} and low magnetic field $\gamma=1$. Left: Spectral Form Factor $g(t,\beta)$. The peak caused by the magnetic field is almost absent (compare to Fig. \ref{fig:tju_hermSFFandPsG36}). Right: Level Spacings Distribution $P(s)$. This model has $\langle r \rangle = 0.5284$. $\langle r \rangle_{GUE} = 0.6027$ and $\langle r \rangle_{GOE} = 0.5359$.}
		\label{fig:tju_hermSFFandPsG1}
\end{figure*}

\begin{figure}
	\centering
	\includegraphics[width=\columnwidth]{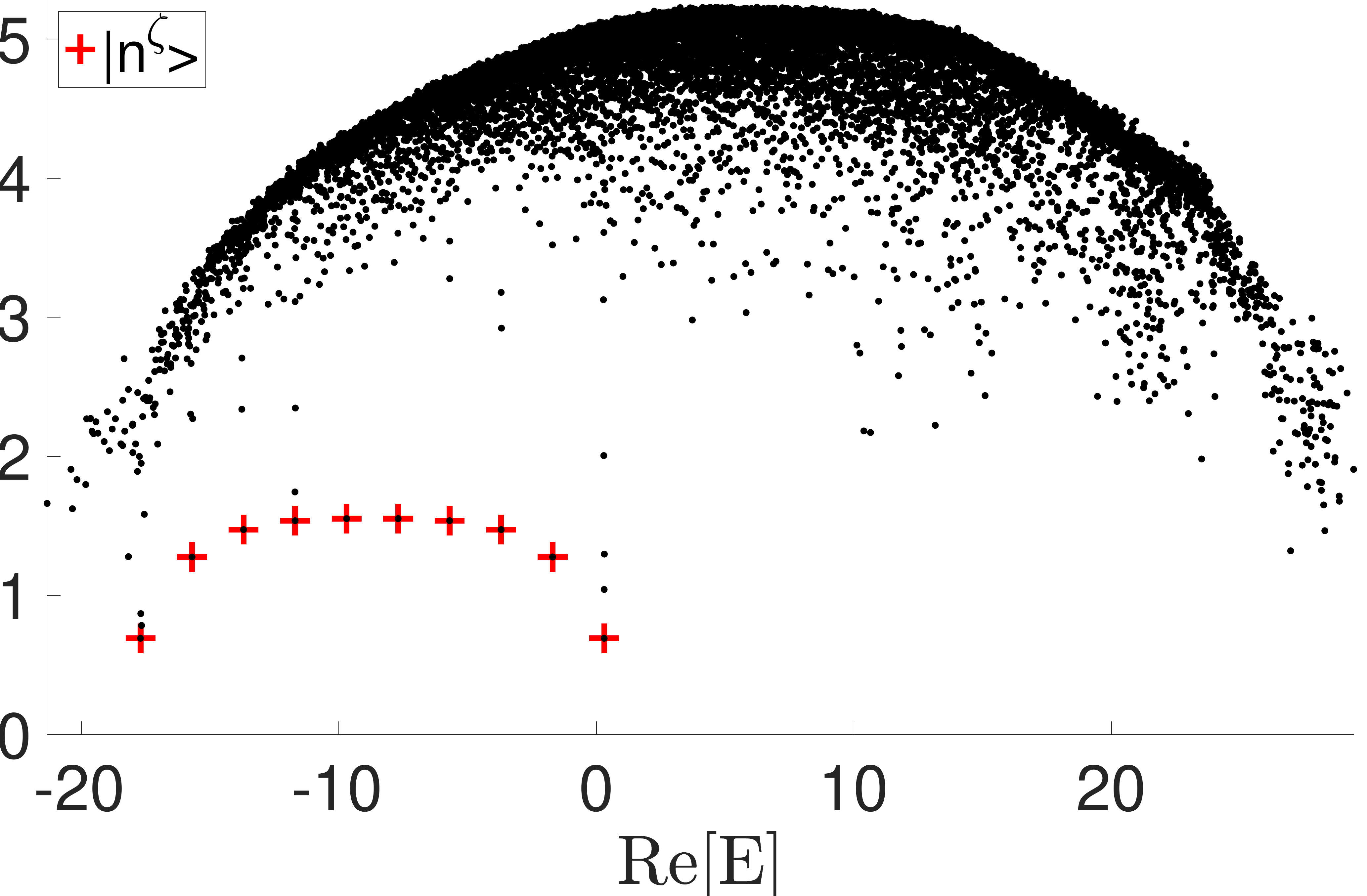}
	\caption{Entanglement entropy for non-Hermitian $tJU$ \eqref{eq:HnumericsNonHerm} in low magnetic field $\gamma=1$.}
		\label{fig:tju_nHermEntrSmallField}
\end{figure}

We can better understand several properties of our system using knowledge from random matrix theory. There are three main random matrix ensembles corresponding to the Hermitian models that we study in this paper; these are commonly known as the Wigner ensembles consisting of the Gaussian Orthogonal Ensemble (GOE), the Gaussian Unitary Ensemble (GUE), and the Gaussian Symplectic Ensemble (GSE). The GOE is time reversal invariant and is a random real symmetric matrix ($H = H^T$ ) where the entries are drawn from a normal Gaussian distribution. The GUE is not time reversal invariant and is a random Hermitian matrix ($H = H^\dagger$) where the entries are drawn from a complex Gaussian distribution. Finally, the GSE is time reversal invariant (but breaks rotational symmetry) and is comprised of real quaternion matrices. Here, we quantify quantum chaos in our models using three distinct measures, and we can compare some of these measures from our model to those of the corresponding random matrix ensemble. We examine the spectral form factor (SFF) \cite{cotler2017black}, the level spacings distribution ($P(s)$), and the mean spacings ratio, $\langle r \rangle$.

\subsubsection{Hermitian Hamiltonian}

The mean level spacings ratio, $\left< r\right>$, is often used to quantify chaos as well as spectral transitions between Wigner-Dyson ensembles. The spacing ratio, $r$, is defined as,
\begin{equation}
r = \frac{\min\left(s_i,s_{i-1}\right)}{\max\left(s_i,s_{i-1}\right)} \quad{\textrm{where}}\quad s_i = \lambda_{i+1}-\lambda_i,
\end{equation}
where $\lambda_i$ is the $i^{\textrm{th}}$ eigenvalue. See equation \eqref{nonherm_ratio} for the definition for a non-Hermitian system. We also see level repulsion in the probability density function of the consecutive level spacings ratio, $P(r)$, as we send $r \rightarrow 0$. The analytic mean level spacings ratios are calculated in \cite{Atas_2013}, and are $\left< r \right> \approx 0.5359$ , $\left< r \right> \approx 0.6027$, and $\left< r \right> \approx 0.6762$ for the GOE, GUE, and GSE respectively. 

Based on the Figs. \ref{fig:tju_hermSFFandPsG36} and \ref{fig:tju_hermSFFandPsG1}, we can conclude that our Hermitian models have a chaotic bulk; the SFF has a dip ramp plateau structure, and the level spacings plots and $\left< r\right>$ values closely match those of the GOE. We note that the peak is absent or much reduced for the moderate magnetic field $\gamma=1$ (see Fig. \ref{fig:tju_hermSFFandPsG1}).

\subsubsection{Non-Hermitian Hamiltonian}
The Ginibre symmetry classes are the non-Hermitian analogs to the Dyson symmetry classes. We can compute the level spacings of our non-Hermitian models and compare to those of the Ginibre random matrix analog. It is also possible to compute complex spacing ratios $\left<r\right>$. For example, the Ginibre GUE (GinUE) $\left<r\right>$ is numerically determined to be $\approx 0.74$ \cite{ProsenPRXNonHermitianR}.

The $r$-value is defined as \cite{ProsenPRXNonHermitianR}.
\begin{equation}
r_n = \left| \frac{E_{n,1} - E_n}{E_{n,2}-E_n} \right| \label{nonherm_ratio}
\end{equation}
where $E_{n,1}$ and $E_{n,2}$ are the nearest  and the next nearest energy levels in the spectrum to the given energy level $E_n$.

We can see from Fig. \ref{fig:tju_nonhermPs}, that our non-Hermitian model shows evidence of a chaotic bulk. The level spacings plot fits closely to that of the Ginibre distribution, and the $\left<r\right>$ value of our model is also close to the Ginibre value. The interpretation of the SFF for the non-Hermitian model is less straightforward, though we do see a dip ramp plateau structure when considering only the real part of the eigenvalues. The dip ramp plateau structure is less clear when considering only the magnitude or imaginary part of the eigenvalues.

\section{Full symmetry of the scar families \label{sec:AppSemi}}
Since the states $\ket{n^\zeta}$ are invariant under the groups U$(N)$  and SU$(2)_{\eta}$, these states are invariant 
under the action of all the commutators. Thus we can take an arbitrary hopping \eqref{hop:T} and commute it with \eqref{hop:eta} and see that they do not commute
\begin{gather}
T^+_{ij} = \sum_\sigma \left[c^\dagger_{i\sigma} c_{j\sigma} + c^\dagger_{j\sigma} c_{i\sigma},\eta^+\right] = \sum_{\sigma,\sigma'} \epsilon_{\sigma\sigma'} c^\dagger_{i\sigma} c^\dagger_{j\sigma'}\ .
\end{gather}
It means that the groups U$(N)$ and SU$(2)_{\eta}$, when joined together, do not form a direct product as in the case of $G_i$'s groups. Rather, they form a so called semi-direct product of these groups. And one can work out that it corresponds to
\begin{gather}
U(N) \triangleleft \rm SU(2)_{\eta} = \rm Sp(2N,\mathbb{C}),
\end{gather}
and the states $\ket{n^\zeta}$ are invariant with respect to this group Sp$(2N,\mathbb{C})$. Moreover, the Hilbert space could be decomposed in the following way
\begin{gather}
G_2 \subset G_{sp}=\rm Sp(2N,\mathbb{C})\times \rm SU(2)_{\rm spin},\notag\\
H = \sum_{k \leq N} \braket{k}_{sp} \otimes \left(n-k\right)_{su},  
\end{gather}
where $\braket{k}_{sp}$ corresponds to an antisymmetric representation of Sp$(2N,\mathbb{C})$ formed by multiplying $k$ copies of the fundamental representation, and $\left(k\right)$ is a usual spin-$k$ representation of SU$(2)_{\eta}$.

The same holds for the groups $\tilde{\rm U}(N)$ and SU$(2)_{\rm spin}$, but then we instead have another $\widetilde{\rm Sp}(2N,\mathbb{C})$ group:
\begin{gather}
G_3 \subset \tilde{G}_{sp}=\widetilde{\rm Sp}(2N,\mathbb{C})\times \rm SU(2)_{\eta}.
\end{gather}

The $\ket{n^\eta}$ states are invariant with respect to $\widetilde{\rm Sp}(2N,\mathbb{C})$. The latter contains $S_i^A$, the local SU$(2)_{spin}$ generators which shows that the Heisenberg interaction is actually of a $TT$ form for the $\ket{n^\eta}$ states. This is another way to see that the Heisenberg interaction exactly annihilates the paired $\ket{n^\eta}$ and $\ket{n^\eta}'$ states.

\bibliography{followUp}
\end{document}